\documentclass[aps,pra,twocolumn,showpacs]{revtex4-2}
\usepackage{epsfig,amssymb,amsfonts}
\usepackage{amsmath,amsfonts,amssymb}
\usepackage{graphicx}
\usepackage{caption}
\usepackage{xcolor}
\input{epsf}
\usepackage{epstopdf}
\usepackage{natbib}

\usepackage[labelformat=simple]{subcaption}
\makeatletter
\renewcommand\p@subfigure{\thefigure}
\makeatother

\begin{document}

\title{Minimum time generation of a uniform superposition in a qubit with only transverse field control}
\author{Vasileios Evangelakos}
\author{Emmanuel Paspalakis}
\author{Dionisis Stefanatos}
\email{dionisis@post.harvard.edu}
\affiliation{Materials Science Department, School of Natural Sciences, University of Patras, Patras 26504, Greece}
\date{\today}
\begin{abstract}
We consider a two-level system with a fixed energy spacing (detuning) between the two levels and a single transverse control field which can take values between zero and a maximum amplitude. Using Pontryagin's maximum principle, we completely solve the problem of generating in minimum time a uniform superposition of the two quantum states when starting from one of them, for all the values of the ratio between the maximum control amplitude and the detuning. For each value of this ratio we find the optimal pulse sequence to have the bang-bang form, and calculate the durations of the pulses composing it. The suggested framework is not only restricted to the problem at hand, but it can be also exploited in the problem of fast charging a quantum battery based on a two-level system, as well as for the optimization of pulse-sequences used for the controlled preparation of the excited state in a quantum emitter, which is a prerequisite for its usage as a single-photon source.

\end{abstract}

\maketitle

\section{Introduction}

Quantum optimal control constitutes a stepping stone to the implementation of many modern quantum technologies \cite{Glaser15,Acin18,StefanatosReview,Koch22}. The successful realization of several essential tasks in quantum computation and quantum sensing, relies heavily on the ability to accurately manipulate the underlying quantum systems using properly designed electromagnetic pulses \cite{Machnes18,Rembold20}. In order to minimize the devastating effect of decoherence and dissipation, it is generally desirable to reach the target quantum states within short times.

Pontryagin's Maximum Principle \cite{Pontryagin62} is the premier mathematical tool for solving optimal control problems and has been also exploited in the context of quantum systems \cite{Boscain21,Venuti21}.
Specifically, this methodology has been successfully used for the time-optimal control of a single two-level system \cite{Boscain06,Bason12,Boscain14,Garon13,Hegerfeldt13,Dionis23}, the simultaneous manipulation of two uncoupled two-level systems \cite{Assemat10,VanDamme18,Ji18}, and even an ensemble of such systems \cite{Martikyan20b}. In all the above works, dissipation and decoherence are ignored. The time-optimal control of a two-level system in the presence of relaxation has been studied in several works \cite{Sugny07,Stefanatos09,Lapert10,Lin20,Lokutsievskiy21}, along with the optimal manipulation of a pair of coupled two-level systems \cite{Khaneja03,Stefanatos04}, mostly in the context of nuclear magnetic resonance. The minimum-time generation of entanglement between coupled two-level systems has been investigated in Ref. \cite{Rahmani18}. The Maximum Principle has been also employed for the efficient population transfer in three-level quantum systems \cite{Sola99,Boscain02,Kumar11,Rat12,Assemat12,Qi19,Dalessandro20,Stefanatos21a,Stefanatos22royal,Evangelakos23,Liu23}, as well as for many-body systems \cite{Chamon17,Stojanovic23}. Other applications include the fast cooling of atoms \cite{Salamon09,Stefanatos11} and condensates \cite{Chamon13}, the fast atomic transport without final excitations \cite{Chen11,Zhang22}, the squeezing of coherent states \cite{Dong15}, the frictionless decompression of condensates \cite{Stefanatos12}, the fast generation of entanglement between coupled harmonic \cite{Stefanatos17} and nonlinear \cite{Stefanatos18} oscillators, and other operations in continuous variable quantum systems \cite{johnsson23}. It has also inspired optimal strategies for the efficient control of a qubit in contact with a structured environment \cite{Fischer19,Ansel22,Pechen23}, for quantum sensing using the electron nuclear spin \cite{Khaneja07,Aiello15}, as well as general numerical methods for searching bang-bang optimal controls \cite{Bukov18,Larocca20,Fei23,bondar23}. A recent more ``exotic" application is the fast charging of quantum batteries \cite{Mazzoncini23}.

In the present work we consider a two-level system without relaxation, with a fixed energy spacing (detuning) between the two levels, corresponding to a fixed longitudinal field, and only available control one transverse field, restricted to take values between zero and a maximum amplitude. Using Maximum Principle and building upon the seminal work of Boscain and Mason \cite{Boscain06}, we completely solve the problem of generating in minimum time a uniform superposition of the two qubit states, when starting from one of them, for all the positive values of the maximum amplitude of the transverse control field. On the Bloch sphere, the problem can be stated as bringing the Bloch vector from initially the north pole to the equator in minimum time, while the problem in Ref. \cite{Boscain06} was to transfer the Bloch vector from the north to the south pole, using a transverse control which could also acquire negative values. We show that when the maximum amplitude is greater or equal to the detuning, then a single On pulse is optimal, where the control takes its maximum value for the whole interval. When the maximum amplitude is smaller than the detuning, the optimal control has the bang-bang form, with alternating On and Off (zero) pulses. For each value of the ratio between the maximum control amplitude and the detuning, we find the candidate optimal pulse-sequences. We show that there are actually two types, one symmetric, where the initial and final On pulses have the same duration, and one complementary, where these durations add up to give the duration of an intermediate On pulse in the sequence. For the second type we find analytical expressions for the durations of the pulses, as functions of the ratio, while for the first type the durations of the pulses are obtained after solving numerically a transcendental equation which depends on that ratio. The comparison between the total durations of the candidate optimal pulse-sequences reveals a simple pattern of optimality. The space between zero and one, where the ratio between the maximum control amplitude and the detuning takes values, is divided into intervals, and within each interval the complementary type is optimal for larger ratio values, but as it decreases and after some point the symmetric type becomes optimal. As we move between consecutive intervals for decreasing ratio, the pulse-sequences acquire an additional On-Off pair of pulses.   

The suggested framework is not only restricted to the problem at hand, but it can be also exploited in the problem of charging a two-level quantum battery in minimum-time \cite{Mazzoncini23}, as well as for the optimization of the SUPER excitation scheme \cite{Super21,Super22}, where a properly modulated square pulse-sequence is used for the controlled preparation of the excited state in a quantum emitter, a prerequisite for its usage as a single-photon source. This article is organized as follows: In Sec. \ref{sec:system} we formulate the optimal control problem, while in Sec. \ref{sec:optimal_when_omega_big} we analyze it and present the solution for maximum control amplitude larger than the detuning. In Sec. \ref{sec:optimal_when_omega_small} we solve the optimal control problem for maximum control amplitude smaller than the detuning, in Sec. \ref{sec:results} we present specific examples, while Sec. \ref{sec:conclusion} concludes this work.

\section{The Optimal Control Problem}
\label{sec:system}

Consider a two-level system with Hamiltonian
\begin{equation}
\label{original_H}
    \hat{H}(t)=\frac{\hbar}{2}\Delta \sigma_{z}+\frac{\hbar}{2}\Omega(t) \sigma_{x}
\end{equation}
where $\sigma_x$, $\sigma_z$ are the Pauli matrices, $\Delta>0$ is the constant detuning and $\Omega(t)\in[0,\Omega_0]$ is the time-dependent Rabi frequency, which is bounded and acts as the control on the system. We would like to find the optimal $\Omega^*(t)$, which drives the qubit from the starting state $\lvert\Psi_0\rangle=\lvert1\rangle$ (up state) to the target uniform superposition state
\begin{equation}
\label{taget_state}
    \lvert\Psi_f\rangle=\frac{1}{\sqrt{2}}\left(\lvert0\rangle+e^{i \varphi}\lvert1\rangle\right),
\end{equation}
with arbitrary phase $\varphi$, in minimum time.
The state of the qubit can be generally expressed as
\begin{equation}
\label{general_state}
    \lvert\Psi(t)\rangle=c_0(t)\lvert0\rangle+c_1(t)\lvert1\rangle,
\end{equation}
so from the Schr\"{o}dinger equation we find the system
\begin{equation}
\label{deq_1}
    i\begin{pmatrix} \dot{c_1} \\ \dot{c_0} \end{pmatrix}=\frac{1}{2} \begin{pmatrix} \Delta & \Omega \\ \Omega & -\Delta \end{pmatrix} \begin{pmatrix} c_1 \\ c_0 \end{pmatrix}.
\end{equation}
The starting state is $c_1(0)=1$, $c_0(0)=0$ and the target state at the final time $t_f$ is such that
\begin{equation}
\label{f_equ}
    \lvert c_1(t_f) \rvert = \lvert c_0(t_f) \rvert =\frac{1}{\sqrt{2}}.
\end{equation}

In order to have real state variables and use the optimal control formalism, we transform the system on the Bloch sphere, using the mapping
\begin{subequations}
\label{set_xyz}
    \begin{eqnarray}
    x &=& c_1 c_0^* + c_0 c_1^* = 2 \Re (c_1 c_0^*), \\
    y &=& i (c_1 c_0^* - c_0 c_1^*) = -2 \Im (c_1 c_0^*), \\
    z &=& \lvert c_1 \rvert^2 - \lvert c_0 \rvert^2 = c_1 c_1^* - c_0 c_0^*. \label{set_z}
    \end{eqnarray}
\end{subequations}
Then, Eq. (\ref{deq_1}) becomes 
\begin{equation}
\label{deq_2}
    \begin{pmatrix} \dot{x} \\ \dot{y} \\ \dot{z} \end{pmatrix}= \begin{pmatrix} 0 & -\Delta & 0 \\ \Delta & 0 & -\Omega \\ 0 & \Omega & 0 \end{pmatrix} \begin{pmatrix} x \\ y \\ z\end{pmatrix},
\end{equation}
which can be written in compact form as
\begin{equation}
\label{Bxr}
    \dot{\vec{r}}=\vec{B}(t) \times \vec{r},
\end{equation}
with the state of the system being expressed by the Bloch vector $\vec{r}=(x,y,z)^T$. This equation corresponds to rotations of the Bloch vector around the instantaneous total field
\begin{equation}
\label{Bfield}
    \vec{B}(t)=\Omega(t)\hat{x}+\Delta\hat{z}.
\end{equation}
The initial state $\vec{r}(0)=(0,0,1)^T$ coincides with the north pole, and the objective is to find the optimal transverse field $\Omega^*(t)\in[0,\Omega_0]$, which brings the Bloch vector on the equator in minimum time. The terminal boundary condition is thus
\begin{equation}
\label{term_z}
  z(t_f)=0  
\end{equation}
and the cost function  to be minimized is
\begin{equation}
\label{run_cost}
    J=\int_{0}^{t_f} 1\,dt.
\end{equation}

Note that a closely related approach to address time-optimal quantum control problems is the \emph{quantum brachistochrone} method \cite{Carlini06}, which also requires the solution of a two-point boundary value problem. As a result, practical solutions of quantum control problems using this method are also relatively rare and typically correspond to highly symmetric cases where the resulting quantum brachistochrone equations can to a large extent be simplified analytically, as in Ref. \cite{Stojanovic22a}. Here we follow the optimal control approach since it is more standard and straightforward, for example we require explicitly the minimization of time and not of a functional involving the quantum state while there is no need for constraints ensuring the preservation of the form of the Hamiltonian. Furthermore, the corresponding algorithms which are necessary for numerical optimal control of more complex quantum systems are well developed.

\section{Analysis of the optimal control problem and solution for $\Omega_0\geq\Delta$}
\label{sec:optimal_when_omega_big}

For the previously stated optimal control problem, the control Hamiltonian is \cite{Heinz12}
\begin{eqnarray}
\label{Hc}
    H_c &=& 1 + \lambda_x\dot{x}+\lambda_y\dot{y}+\lambda_z\dot{z} \nonumber\\
        &=& 1 + \lambda_x(-\Delta y) + \lambda_y(\Delta x - \Omega z) + \lambda_z \Omega y \nonumber  \\
        &=& 1 + \left( \lambda_y x-\lambda_x y \right)\Delta +\left( \lambda_z y-\lambda_y z \right)\Omega,
\end{eqnarray}
where the first terms comes from the integrand of the running cost (\ref{run_cost}) while the costates $\lambda_x,\lambda_y,\lambda_z$ satisfy the adjoint equations
\begin{subequations}
\label{Hc}
\begin{eqnarray}
    \dot{\lambda}_x &=& - \frac{\partial H_c}{\partial x} = -\Delta \lambda_y, \\[5pt]
    \dot{\lambda}_y &=& - \frac{\partial H_c}{\partial y} = \Delta \lambda_x - \Omega \lambda_z, \\[5pt]
    \dot{\lambda}_z &=& - \frac{\partial H_c}{\partial z} = \Omega \lambda_y.
\end{eqnarray}
\end{subequations}
These can be written in compact form as
\begin{equation}
\label{Bxlambda}
    \dot{\vec{\lambda}}=\vec{B}(t) \times \vec{\lambda},
\end{equation}
where $\vec{\lambda}=(\lambda_x,\lambda_y,\lambda_z)^T$, and are the same with the state equations (\ref{Bxr}) for the Bloch vector. The terminal boundary conditions for the costates can be found from the general expression \cite{Heinz12}
\begin{equation}
\label{lambda_tf}
    \lambda^T(t_f)=\frac{\partial \Phi}{\partial r(t_f)} + v \frac{\partial \phi}{\partial r(t_f)},
\end{equation}
where $\Phi=\Phi(r(t_f),t_f)$ is the terminal cost and $\phi=\phi(r(t_f),t_f)$ is the terminal boundary condition, with $\Phi=0$ and $\phi=z(t_f)$ in this particular case, while $v$ is a Langrange multiplier to be determined. Consequently
\begin{subequations}
\label{lambdas_tf}
    \begin{eqnarray}
    \lambda_x(t_f) &=& v \frac{\partial z(t_f)}{\partial x(t_f)} = 0, \\[5pt]
    \lambda_y(t_f) &=& v \frac{\partial z(t_f)}{\partial y(t_f)} = 0, \\[5pt]
    \lambda_z(t_f) &=& v \frac{\partial z(t_f)}{\partial z(t_f)} = v.
    \end{eqnarray}
\end{subequations}

Using the above relations we can show that the inner product of state and costate vectors is constant and equal to zero
\begin{equation}
\label{lamdba}
    \vec{\lambda}\cdot\vec{r}=0.
\end{equation}
Indeed,
\begin{eqnarray}
\label{der_l.r}
    \frac{d}{dt}(\vec{\lambda}\cdot\vec{r}) &=& \dot{\vec{\lambda}}\cdot\vec{r} + \vec{\lambda}\cdot\dot{\vec{r}} \nonumber\\[5pt]
    &=& \left( \vec{B} \times \vec{\lambda} \right) \cdot\vec{r} + \vec{\lambda}\cdot \left( \vec{B} \times \vec{r} \right)  \nonumber\\[5pt]
    &=& \vec{\lambda}\cdot \left( \vec{r} \times \vec{B}\right) + \vec{\lambda}\cdot \left( \vec{B} \times \vec{r} \right) \nonumber\\[5pt]
    &=& 0,
\end{eqnarray}
and
\begin{equation}
\label{lamdba}
    \vec{\lambda}(t_f)\cdot\vec{r}(t_f)=\lambda_x(t_f) x(t_f) + \lambda_y(t_f) y(t_f) + \lambda_z(t_f) z(t_f) = 0.
\end{equation}
From this we can find using the initial conditions at t=0 that $\lambda_z(0)=0$. 

We next move to find the optimal pulse sequence. The control Hamiltonian can be expressed as
\begin{equation}
\label{Hc_phi}
    H_c = 1 + \phi_x\Omega + \phi_z \Delta,
\end{equation}
where
\begin{subequations}
\label{setting_phis}
    \begin{eqnarray}
    \phi_x &=& y\lambda_z - z\lambda_y, \\
    \phi_y &=& -x\lambda_z + z\lambda_x, \\
    \phi_z &=& x\lambda_y - y\lambda_x, \label{setting_phis_z}
    \end{eqnarray}
\end{subequations}
According to Pontryagin's Maximum Principle \cite{Heinz12}, the optimal control $\Omega^*(t)$ is selected to minimize the control Hamiltonian for almost all times (except probably a measure zero set), while the control Hamiltonian for the optimal $\Omega^*(t)$ is constant, equal to zero for minimum time problems like the one studied here, $H_c=0$. Since the control Hamiltonian is linear in the bounded variable $\Omega\in[0,\Omega_0]$, the optimal pulse sequence is determined by the switching function $\phi_x$ multiplying the control $\Omega(t)$. Specifically, $\Omega^*(t)=0$ while $\phi_x>0$ and $\Omega^*(t)=\Omega_0$ while $\phi_x<0$. The Maximum Principle provides no information about the optimal control for finite time intervals where the switching function is zero. In such cases the control is called singular and is determined from the requirements $\phi_x=\dot{\phi}_x=\ddot{\phi}_x=\ldots=0$. It becomes obvious that, in order to find the optimal $\Omega^*(t)$, it is necessary to track the time evolution of $\phi_x$. Using the state and costate equations, it is not hard to verify that the components of the vector
\begin{equation}
\label{phi_as_cross}
    \vec{\phi} = \left( \phi_x,\phi_y,\phi_z \right)^T = \vec{r} \times \vec{\lambda}
\end{equation}
satisfy the equation
\begin{subequations}
\label{phi_dots}
    \begin{eqnarray}
    \dot{\phi}_x &=& -\Delta \phi_y, \label{phi_dots_x} \\[5pt]
    \dot{\phi}_y &=& \Delta \phi_x - \Omega \phi_z, \label{phi_dots_y} \\[5pt]
    \dot{\phi}_z &=& \Omega \phi_y, \label{phi_dots_z}
    \end{eqnarray}
\end{subequations}
thus also
\begin{equation}
\label{Bxphi}
    \dot{\vec{\phi}}=\vec{B}(t) \times \vec{\phi}.
\end{equation}

Now we examine the possibility for singular controls. If we assume that $\phi_x=0$ in a finite time interval then, for $\Delta\neq0$, we also get $\phi_y=0$ from Eq. (\ref{phi_dots_x}) and $\dot{\phi_z}=0$ from Eq. (\ref{phi_dots_z}). The constant value of $\phi_z$ is found as $\phi_z(t)=-1/\Delta$, from Eq. (\ref{Hc_phi}) and the requirement of Maximum Principle that $H_c=0$ at all times for a minimum time problem. From the above analysis and Eq. (\ref{phi_dots_y}), on the singular arc we also have
\begin{equation}
\label{onsing}
\begin{array}{ccccc}
    \dot{\phi_y}=0 &\Rightarrow& \Omega\phi_z=0 &\Rightarrow& \Omega=0  
\end{array}
\end{equation}
Thus, on the singular arc the control field is zero. Using Eq. (\ref{phi_as_cross}) we can show that the inner product $\vec{\phi}\cdot\vec{r}$ is constant, with value
\begin{eqnarray}
\label{phidotr}
    \vec{\phi}\cdot\vec{r}&=& \vec{\phi}(0)\cdot\vec{r}(0) \nonumber\\
                          &=&\phi_x(0)x(0)+\phi_y(0)y(0)+\phi_z(0)z(0) \nonumber\\
                          &=& 0,
\end{eqnarray}
where we have used that $x(0)=y(0)=0$ and $\phi_z(0)=0$ from Eq. (\ref{setting_phis_z}).
On the singular arc, where $\phi_x=\phi_y=0$ and $\phi_z=-1/\Delta\neq0$, the above relation gives that $z=0$. Consequently, singular arcs are contained in the equator and are thus not encountered in the optimal solution of the present problem, where the target is to reach the equator in minimum time. 

The optimal control switches between the boundary values $0$ and $\Omega_0$, having the so called bang-bang form. Because $\phi_z(0)=0$, the optimal pulse sequence should start with $\Omega(t)=\Omega_0$, to avoid $H_c(0)=1$. Also, it should end with $\Omega(t)=\Omega_0$, since $\Omega(t)=0$ corresponds just to a rotation around z-axis. We next show that an Off bang, $\Omega(t)=0$, cannot occur in between for the case where $\Omega_0>\Delta$. Let $t_s$ be a switching time from $\Omega(t)=\Omega_0$ to $\Omega(t)=0$, thus
\begin{equation}
\label{swithtime}
\phi_x(t_s)=0 \text{  and  } \dot{\phi_x}(t_s)=A>0  
\end{equation}
From Eqs. (\ref{phi_dots}) we find that, during the off bang, the switching function obeys the equation
\begin{equation}
\label{off_equ}
    \ddot{\phi_x} = -\Delta \dot{\phi_y} = - \Delta^2 \phi_x. 
\end{equation}
For the initial conditions (\ref{swithtime}) at $t=t_s$,
the solution is
\begin{equation}
\label{phix_off}
    \phi_x(t) = A \sin{\left[\Delta(t-t_s)\right]}.
\end{equation}
The duration of the Off bang is the time it takes for $\phi_x$ to become zero again, thus
\begin{equation}
\label{off_dur}
    t_{\text{off}} = \frac{\pi}{\Delta}  
\end{equation}
A pulse sequence containing such a bang should last longer than the Off duration, $t_f > t_{\text{off}}$.

We next show that when $\Omega_0>\Delta$ a single On pulse $\Omega(t)=\Omega_0$ can reach the target in shorter time than Eq. (\ref{off_dur}). For easiness we will perform the calculations using the corresponding constant quantum mechanical Hamiltonian
\begin{eqnarray}
\label{H_n}
    \hat{H}&=&\frac{\hbar}{2}(\Delta \sigma_{z}+\Omega_0 \sigma_{x}) \nonumber\\
              &=&\frac{\hbar \sqrt{\Delta^2+\Omega_0^2}}{2} \left( n_x \sigma_x + n_z \sigma_{z} \right),
\end{eqnarray}
where
\begin{equation}
\label{n_def}
\begin{array}{ccc}
    n_x &=& \frac{\Omega_0}{\sqrt{\Delta^2+\Omega_0^2}}, \\
    n_z &=& \frac{\Delta}{\sqrt{\Delta^2+\Omega_0^2}},
\end{array}
\end{equation}
with $n_x^2+n_z^2=1$.
The corresponding propagator is
\begin{equation}
\label{prop_on}
    U(t_f)=e^{-\frac{i}{\hbar} \hat{H} t_f} = e^{-\frac{i}{2} \gamma \hat{n} \cdot \hat{\sigma}},
\end{equation}
with $\gamma = t_f\sqrt{\Delta^2+\Omega_0^2}$ and $\hat{n} = (n_x,0,n_z)^T$. Using the following well known identity involving Pauli matrices
\begin{equation}
\label{identity}
    e^{-\frac{i}{2} \gamma \hat{n} \cdot \hat{\sigma}} = \cos \left(\frac{\gamma}{2}\right)  \hat{I} - i \sin \left(\frac{\gamma}{2}\right)  (\hat{n} \cdot \hat{\sigma}),
\end{equation}
we get
\begin{equation}
\label{prop_on_n}
    U(t_f)= \begin{bmatrix} \cos{\frac{\gamma}{2}} - i n_z \sin{\frac{\gamma}{2}} & - i n_x \sin{\frac{\gamma}{2}} \\  - i n_x \sin{\frac{\gamma}{2}} & \cos{\frac{\gamma}{2}} + i n_z \sin{\frac{\gamma}{2}} \end{bmatrix}.
\end{equation}
From the relation
\begin{equation}
\label{final_state}
    \lvert\Psi(t_f)\rangle=U(t_f) \lvert\Psi_0\rangle
\end{equation}
and since the initial state is $\lvert\Psi_0\rangle=(1,0)^T$, we find the following probability amplitudes for the final state
\begin{subequations}
\label{c1_c0_final}
    \begin{eqnarray}
    c_1(t_f) &=& \cos \frac{\gamma}{2}-i n_z \sin \frac{\gamma}{2}, \\
    c_0(t_f) &=& -i n_x \sin \frac{\gamma}{2}.
    \end{eqnarray}
\end{subequations}
The terminal condition that the final point lies on the equator, $z(t_f)=|c_1(t_f)|^2-|c_0(t_f)|^2=0$, gives
\begin{equation}
\label{final_condition}
    \cos^2 \frac{\gamma}{2} + n_z^2 \sin^2 \frac{\gamma}{2} = n_x^2 \sin^2 \frac{\gamma}{2},
\end{equation}
and using that $n_z^2 = 1-n_x^2$ we find
\begin{equation}
\label{gamma}
\sin \frac{\gamma}{2}=\frac{1}{\sqrt{2}n_x}=\sqrt{\frac{1}{2}\left[1+\left(\frac{\Delta}{\Omega_0}\right)^2\right]}.
\end{equation}
For $\Omega_0>\Delta$ the right hand side of the above equation is smaller than one, thus there is a solution $\gamma/2<\pi/2\Rightarrow \gamma<\pi$. But $\gamma = t_f\sqrt{\Delta^2+\Omega_0^2}$, so we end up with the relation
\begin{equation}
\label{final_time_constraint}
    t_f < \frac{\pi}{\sqrt{\Delta^2+\Omega_0^2}} < \frac{\pi}{\Delta}
\end{equation}
From this inequality we conclude that for $\Omega_0>\Delta$ the single On pulse reaching the target is shorter than any pulse sequence containing Off pulses, thus it is the optimal solution.

\begin{figure}[t]
 \centering
 \includegraphics[width=\linewidth]{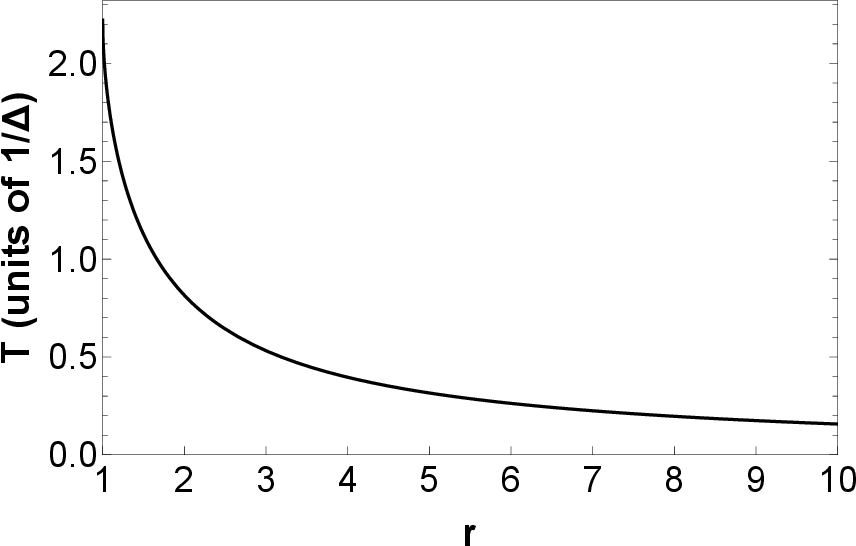}
\caption{Duration of the optimal single On pulse versus $r=\Omega_0/\Delta$, for $r\geq 1$.}
\label{fig:durations_for_r_1}
\end{figure}

The duration of the pulse is determined by $\Omega_0$ and $\Delta$, which also determine the terminal position on the equator and the relative phase $\phi$ of the superposition. If we define parameter $r$ as the ratio
\begin{equation}
\label{r}    
r=\frac{\Omega_0}{\Delta},
\end{equation} 
then from Eq. (\ref{gamma}) we get 
\begin{equation}
\label{T_one_on}
t_f\Delta=\frac{2}{\sqrt{r^2+1}}\sin^{-1}\sqrt{\frac{1}{2}\left(1+\frac{1}{r^2}\right)}.
\end{equation}
In Fig. \ref{fig:durations_for_r_1} we plot the duration of the optimal pulse versus $r$.

\section{Optimal Solution when $\Omega_0<\Delta$}
\label{sec:optimal_when_omega_small}

When $\Omega_0<\Delta$, the right hand side of Eq. (\ref{gamma}) is larger than one, so there is no real $\gamma$ to satisfy it and a single On pulse cannot bring the Bloch vector on the equator. In this case, the optimal solution consists of a sequence of On and Off pulses, since no singular control is encountered as it was shown in the previous section. Additionally, bacause an Off pulse corresponds to a rotation around $z$-axis, the optimal pulse sequence for the problem under investigation cannot end with such a pulse, but has the form On-Off-On-$\ldots$-Off-On. The number and duration of these pulses depend on the ratio $r=\Omega_0/\Delta$.
In this section we will determine the corresponding optimal pulse-sequences. 

The switching function $\phi_x$ determines the duration of the individual On and Off pulses, as explained in the previous section. We can obtain an equation for $\phi_x$ working as follows. We take the time derivative in Eq. (\ref{phi_dots_x}) and substitute $\dot{\phi}_y$ from Eq. (\ref{phi_dots_y}) to find
\begin{equation}
\label{phi_x_ddot}
    \ddot{\phi}_x=-\Delta \dot{\phi}_y = -\Delta^2 \phi_x + \Delta \Omega \phi_z.
\end{equation}
For a minimum-time optimal control problem, as we have here, the control Hamiltonian is zero throughout $H_c=0$ according to optimal control theory \cite{Heinz12}. This condition, through Eq. (\ref{Hc_phi}), is translated to the relation
\begin{equation}
\label{phi_z}
    \phi_z=-\frac{1+\phi_x \Omega}{\Delta}.
\end{equation}
Substituting the above expression for $\phi_z$ in Eq. (\ref{phi_x_ddot}) we get
\begin{equation}
\label{phi_x_deq}
    \ddot{\phi}_x + \left( \Omega^2 + \Delta^2 \right) \phi_x + \Omega = 0,
\end{equation}
where note that during each individual pulse the control $\Omega(t)$ is held constant, equal to $\Omega_0$ for On pulses and $0$ for Off pulses.

Solving this second order differential equation during the On and Off pulses will provide insight regarding their duration. Observe that it is easier to solve Eq. (\ref{phi_x_deq}) in each time interval where $\Omega(t)$ is held constant. We will use the following notation to distinguish the solution in each domain 
\begin{equation}
\phi_x=
    \begin{cases}
        \phi_x^{(\text{I})}(t) & \text{if } t \in \text{first On}, \\
        \phi_x^{(\text{II})}(t) & \text{if } t \in \text{Off}, \\
        \phi_x^{(\text{III})}(t) & \text{if } t \in \text{subsequent On}.
    \end{cases}
\end{equation}
Note that we treat separately the cases of the initial and subsequent On pulses, since the corresponding boundary conditions for $\phi_x$ are different, as we will explain below.
For simplification of the results we are also defining a modified time as
\begin{equation}
\label{mod_time}
    \tilde{t} = t \sqrt{\Delta ^2+\Omega ^2}.
\end{equation}

\begin{figure}[h]
 \centering
 \begin{subfigure}[b]{0.4\textwidth}
    \centering\caption{}\includegraphics[width=\linewidth]{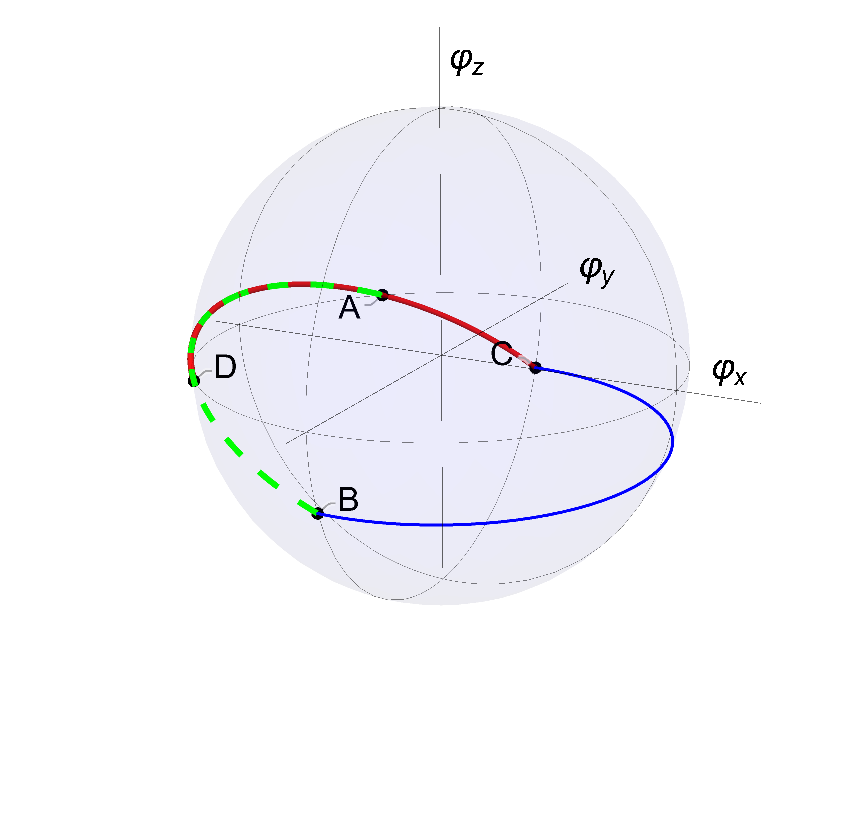}\label{fig:fx_1}
\end{subfigure}
\hspace{.2cm}
\begin{subfigure}[b]{0.4\textwidth}
    \centering\caption{}\includegraphics[width=\linewidth]{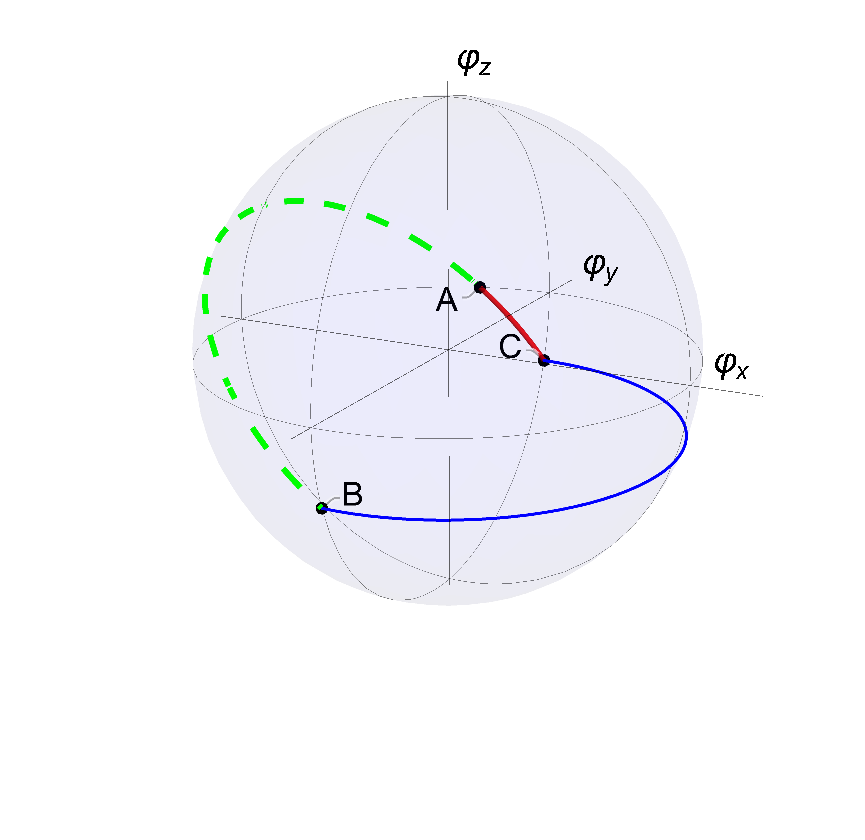}\label{fig:fx_2}
\end{subfigure}
\caption{Trajectory of $\vec{\phi}$ in the cases where the first and the last On pulses are (a) Equal and (b) Complementary}
\label{fig:fx}
\end{figure}

Let $s$ denote the duration of the first On pulse. The boundary conditions for this time interval are $\phi^{(\text{I})}_x(0)=-1/\Omega_0$, which rises from the fact that $H_c=0$ and $\phi_z(0)=0$, as we discussed in the previous section, and $\phi^{(\text{I})}_x(s)=0$ because the On pulse terminates when the switching function becomes zero for the first time. Using these boundary conditions we find
\begin{equation}
\label{phi_x_1on}
    \phi^{(\text{I})}_x(t) = \frac{\frac{\sin \tilde{t}}{\sin \tilde{s} } \left(\cos \tilde{s}+r^2\right)-\cos \tilde{t}-r^2}{\Delta  r \left(r^2+1\right)}.
\end{equation}
As displayed in Fig. \ref{fig:fx}, during the first On pulse the vector $\vec{\phi}$ travels the dashed green trajectory from point A on the $\phi_z=0$ plane to point B on the $\phi_x=0$ plane.

We next move to the first Off pulse. Since we are interested in the duration of the pulse, we will set the zero of time at the beginning of the pulse and solve Eq. (\ref{phi_x_deq}) with this convention. The boundary conditions now are that the switching function should vanish at the beginning and end. But instead of using the terminal condition, we can exploit that $\dot{\phi}_x=-\Delta \phi_y$ is continuous function, because $\phi_y$ is continuous, and use the initial value of $\dot{\phi}_x$ instead. Thus, the initial conditions for the first Off pulse are $\phi^{(\text{II})}_x(0)=\phi^{(\text{I})}_x(s)=0$ and
\begin{equation}
\label{dot_phi_x_at_s}
    \dot{\phi}^{(\text{II})}_x(0)=\dot{\phi}^{(\text{I})}_x(s) = \frac{ r^2 \cot \tilde{s} + \csc \tilde{s} }{ r \sqrt{r^2+1} }
\end{equation}
Solving Eq. (\ref{phi_x_deq}) with the above conditions we get that 
\begin{equation}
\label{phi_x_off}
    \phi^{(\text{II})}_x(t) = \frac{r^2 \cot \tilde{s}+\csc \tilde{s}}{\Delta  r \sqrt{r^2+1}} \sin (\Delta  t),
\end{equation}
which of course has the form (\ref{phix_off}). From the above relation it is obvious that the duration of the Off pulse is $t_{\text{off}}=\pi/\Delta$, as given in Eq. (\ref{off_dur}), because  $\phi^{(\text{II})}_x(t_{\text{off}})=0$ for the first time after the onset of this pulse. During the Off pulse, vector $\vec{\phi}$ travels the solid blue trajectory, counterclockwise between points B and C of the $\phi_x=0$ plane.  

Working similarly as before and using the duration of the Off pulse, we can find the initial conditions for the next On pulse. We have $\phi^{(\text{III})}_x(0)=\phi^{(\text{II})}_x(t_{\text{off}})=0$ and
\begin{equation}
\label{dot_phi_x_at_toff}
   \dot{\phi}^{(\text{III})}_x(0)=\dot{\phi}^{(\text{II})}_x(t_{\text{off}}) =-\frac{r^2 \cos \tilde{s}+1}{r \sqrt{r^2+1} \sin \tilde{s}}
\end{equation}
Solving Eq. (\ref{phi_x_deq}) for the On pulse with those initial conditions we get
\begin{eqnarray}
\label{phi_x_ron}
    \phi^{(\text{III})}_x(t) &=& \frac{r^2 \left[\cos \tilde{t} - \cot \tilde{s} \sin \tilde{t} -1 \right] - \csc \tilde{s} \sin \tilde{t}}{\Delta  r\left( r^2+1 \right)} \nonumber \\
    &=& -\frac{r}{\Delta(r^2+1)}\left[1+\frac{\sin{(\tilde{t}-\theta)}}{\sin\theta}\right],
\end{eqnarray}
where angle $\theta$ is
\begin{equation}
\label{theta}
\theta=\tan ^{-1}\left( \frac{r^2 \sin \tilde{s}}{1+r^2 \cos \tilde{s}} \right).
\end{equation}
If the On pulse is not the last in the sequence, then its duration $t_{\text{on}}$ is such that condition $\phi^{(\text{III})}_x(t_{\text{on}})=0$ is satisfied for the first time from the onset of the pulse. From Eq. (\ref{phi_x_ron}) we easily get
\begin{equation}
\label{t_on(s)}
    \tilde{t}_{\text{on}} = \pi+2\theta= \pi +2 \tan ^{-1}\left( \frac{r^2 \sin \tilde{s}}{1+r^2 \cos \tilde{s}} \right),
\end{equation}
where $t_{\text{on}}$ is expressed as a function of the duration $s$ of the first On pulse, while note that the longer solution $\tilde{t}_{\text{on}}=2\pi$, corresponding to a whole period, is rejected. Vector $\vec{\phi}$ travels the inclined trajectory shown in Fig. \ref{fig:fx}, counterclockwise between points C and B. 

For the subsequent Off and On pulses which may exist in the pulse-sequence, the boundary conditions are the same as described in the corresponding cases above, so we repeatedly get the same solutions $\phi^{(\text{II})}_x(t)$ and $\phi^{(\text{III})}_x(t)$, respectively. During the Off pulses, vector $\vec{\phi}$ traces the horizontal blue trajectory from B to C in time $t_{\text{off}}=\pi/\Delta$, while during the intermediate On pulses (between the first and the last) traces the inclined trajectory between C and B in time $t_{\text{on}}$, which is expressed in terms of the duration $s$ of the first On pulse through Eq. (\ref{t_on(s)}).




The only thing left is to find an expression for the duration $f$ of the final On pulse. From the terminal conditions for the costates (\ref{lambdas_tf}) and the definition (\ref{setting_phis}) we obtain $\phi_z(f)=0$ at the terminal point. This relation, in combination with Eq. (\ref{phi_z}), leads to the final condition $\phi^{(\text{III})}_x(f)=-1/\Omega_0$. Observe that the terminal conditions for $\phi_x, \phi_z$ are the same as the initial conditions for the first On pulse. This symmetry leads to the two possible situations depicted in Fig. \ref{fig:fx}. In the symmetric case, shown in Fig. \ref{fig:fx_1}, vector $\vec{\phi}$ travels the red segment from C to D, where point D on the $\phi_z=0$ plane is symmetric to point A with respect the $\phi_x$-axis. Because of this symmetry, the duration of the last On pulse (red segment) is equal to that of the first On pulse (green dashed segment). In the complementary case, shown in Fig. \ref{fig:fx_2}, vector $\vec{\phi}$ returns to point A from point C along the red segment. Obviously, now the sum of the durations of the first and last On pulses equals the duration of the intermediate On pulses. We thus have the following two cases
\begin{equation}
\label{finalon}
f=
\begin{cases}
    s & \text{symmetric case}, \\
    t_{\text{on}} - s & \text{complementary case}.
\end{cases}
\end{equation}
The above geometric considerations can be easily verified algebraically. For the symmetric case, using the first line of Eq. (\ref{phi_x_ron}) one can immediately see that $\phi^{(\text{III})}_x(s)=-1/\Omega_0$. Also, for the complementary case it is $\sin(\tilde{t}_{\text{on}}-s-\theta)=\sin(\pi+\theta-s)=\sin(s-\theta)$, and from the second line of Eq. (\ref{phi_x_ron}) we get $\phi^{(\text{III})}_x(t_{\text{on}} - s)=\phi^{(\text{III})}_x(s)=-1/\Omega_0$.

We have expressed the durations of all pulses as functions of the duration of the initial On pulse $s$. The value of $s$ will be determined from the condition that the final point of the trajectory should lie on the equator. In order to apply this condition, we need to find the total propagator corresponding to the symmetric and complementary cases. Since the optimal pulse-sequences in both cases include concatenations of Off and full (intermediate) On pulses, it is useful to find the propagator corresponding to a sequence of $n$ such On-Off pairs. Using identity (\ref{identity}) we find the propagators for the single full On and Off pulses,
\begin{equation}
\label{propagator_off_general}
    U_{\text{off}} = -i\sigma_z, 
\end{equation}
\begin{equation}
\label{propagator_on_general}
    U_{\text{on}} = \cos \frac{\tilde{t}_{\text{on}}}{2}  \hat{I} - \frac{i \sin \frac{\tilde{t}_{\text{on}}}{2}}{\sqrt{r ^2+1}}  \left( r \sigma_x +\sigma_z  \right).
\end{equation}
The propagator corresponding to a pair of an Off pulse followed by a full On pulse can be expressed as a single rotation around a different axis as
\begin{equation}
\label{propagator_onoff_exponential}
    U_{\text{on}} U_{\text{off}} = -e^{-\frac{i}{2} \alpha \hat{\nu}\cdot \hat{\sigma}},
\end{equation}
where the angle $\alpha$ and unit vector $\hat{\nu} = (0,\nu_y,\nu_z)^T$ are defined as
\begin{subequations}
\label{nu_yz_r}
    \begin{eqnarray}
    \alpha &=& 2 \cos ^{-1}\left(\frac{\sin \frac{\tilde{t}_{\text{on}}}{2}}{\sqrt{r^2+1}}\right), \label{alpha}\\[10pt]
    \nu_y &=& \frac{ r \sin \frac{\tilde{t}_{\text{on}}}{2}}{\sqrt{r^2+1} \sin \frac{\alpha}{2}}, \\[10pt]
    \nu_z &=& -\frac{ \cos \frac{\tilde{t}_{\text{on}}}{2}}{ \sin \frac{\alpha}{2}}.
    \end{eqnarray}
\end{subequations}
Note that both $\alpha$ and $\hat{\nu}$ are functions of the duration $s$ of the first On pulse.
The propagator corresponding to a sequence of $n$ such On-Off pairs is thus
\begin{equation}
\label{propagator_onoff_exponential}
    \left( U_{\text{on}}U_{\text{off}} \right)^n = (-1)^n  e^{-\frac{i}{2} n \alpha \hat{\nu}\cdot \hat{\sigma}}.
\end{equation}
This relation will become handy in the following.

\subsection{Complementary Case}

The total propagator for a candidate optimal pulse-sequence containing $n$ Off pulses and for the case where the first and the last On pulses are complementary, thus their durations $s, s^*$ satisfy $s+s^*=t_\text{on}$, is
\begin{eqnarray}
\label{total_prop_compl}
    U &=& U_{s^*}  U_{\text{off}}  \underbrace {U_{\text{on}} U_{\text{off}}\cdots U_{\text{on}} U_{\text{off}} }_{n-1}  U_{s} \nonumber
    \\
   \Rightarrow U_{s}  U &=& \underbrace{U_{s}U_{s^*}}_{\text{on}} U_{\text{off}}  \underbrace {U_{\text{on}} U_{\text{off}}\cdots U_{\text{on}} U_{\text{off}} }_{n-1}  U_{s} \nonumber
    \\
   \Rightarrow U_{s}  U &=& \underbrace {U_{\text{on}} U_{\text{off}}\cdots U_{\text{on}} U_{\text{off}} }_{n}  U_{s} \nonumber
    \\
   \Rightarrow U_{s}  U &=& (-1)^{n}  e^{-\frac{i}{2} n \alpha \hat{\nu}\cdot \hat{\sigma}}  U_{s} \nonumber
    \\
   \Rightarrow U &=& (-1)^{n}  U_{-s}  e^{-\frac{i}{2} n \alpha \hat{\nu}\cdot \hat{\sigma}}  U_{s},
\end{eqnarray}
where $\alpha, \hat{\nu}$ are given in Eq. (\ref{nu_yz_r}).
Using identity (\ref{identity}) we find that
\begin{equation}
\label{t_propagator}
    U =  (-1)^{n} e^{-\frac{i}{2} \beta_c \hat{\mu}_c\cdot \hat{\sigma}}, 
\end{equation}
where
\begin{equation}
\beta_c=n\alpha
\end{equation}
is the rotation angle
and $\hat{\mu}_c = (\mu_{cx},\mu_{cy},\mu_{cz})^T$ is the rotation axis corresponding to the total propagator, with
\begin{subequations}
\label{muc_xz}
\begin{eqnarray}
    \mu_{cx} &=& \frac{\sqrt{r^2+1} \nu _y \sin \tilde{s}-r \nu _z \left(\cos \tilde{s}-1\right)}{r^2+1} \nonumber\\
                    &=& \frac{2 r \sin \frac{\tilde{s}}{2} \sin \left(\frac{\tilde{t}_{\text{on}}-\tilde{s}}{2}\right)}{(r^2+1)\sin{\frac{\alpha}{2}}},
                \label{mucx}\\[10pt]
    \mu_{cy} &=& \frac{\left(r^2+1\right) \nu _y \cos \tilde{s}+r \sqrt{r^2+1} \nu _z \sin \tilde{s}}{r^2+1} \nonumber\\
                    &=&  \frac{r \sin \left(\frac{\tilde{t}_{\text{on}}}{2}-\tilde{s}\right)}{\sqrt{r^2+1}\sin{\frac{\alpha}{2}}}, \label{mucy}\\[10pt]
    \mu_{cz} &=& \frac{\nu _z \left(r^2 \cos \tilde{s}+1\right)-r \sqrt{r^2+1} \nu _y \sin \tilde{s}}{r^2+1} \nonumber\\
                &=& 0 \label{mucz}.
\end{eqnarray}
\end{subequations}
Note that the second lines of Eqs. (\ref{muc_xz}) are obtained using Eqs. (\ref{nu_yz_r}), (\ref{t_on(s)}).

Since the rotation axis $\hat{\mu}_c$ corresponding to the total propagator lies on the $xy$-plane, the rotation angle should be $\beta_c=\pi/2$, for the Bloch vector to end up on the equator. We proceed to derive this algebraically. Using propagator (\ref{t_propagator}) we find the following probability amplitudes of the final state,
\begin{eqnarray}
\label{c1_c0_final}
    c_1(t_f) &=& (-1)^{n}\cos\frac{\beta_c}{2}, \\
    c_0(t_f) &=& (-1)^{n}\sin\frac{\beta_c}{2}(\mu _{cy}-i\mu _{cx}).
\end{eqnarray}
The final requirement $z(t_f)= \lvert c_1(t_f) \rvert^2 - \lvert c_0(t_f)\rvert^2$=0 leads to the condition
\begin{eqnarray}
\cos^2\frac{\beta_c}{2}-\sin^2\frac{\beta_c}{2}(\lvert\mu _{cy}\rvert^2+\lvert\mu _{cx}\rvert^2) &=& 0 \nonumber\\
\Rightarrow\cos^2\frac{\beta_c}{2}-\sin^2\frac{\beta_c}{2} &=& 0 \nonumber\\
\Rightarrow\cos\beta_c &=& 0,
\end{eqnarray}
which implies that
\begin{equation}
\label{alpha_in_equal}
     \beta_c=n\alpha = \pi/2,
\end{equation}
as previously anticipated using geometric arguments. The final state on the Bloch sphere is
easily found to be
\begin{equation}
\label{c_axis}
    \vec{r}(t_f) = \left(\mu_{cy}, -\mu_{cx}, 0 \right)^T.
\end{equation}

We finally find the durations of the pulses, which also determine $\mu_{cx}, \mu_{cy}$. Using expression (\ref{alpha}) in Eq. (\ref{alpha_in_equal}) we find the duration of the intermediate (full) On pulses, as a function of $r=\Omega_0/\Delta$,
\begin{equation}
\label{t_on_complementary}
\tilde{t}_{\text{on}} = 2 \pi - 2\sin ^{-1}\left(\sqrt{r^2+1} \cos \frac{\pi }{4 n}\right),
\end{equation}
where note that $\pi< \tilde{t}_{\text{on}} <2\pi$ from Eq. (\ref{t_on(s)}).
Using expression (\ref{t_on_complementary}) in Eq. (\ref{t_on(s)}) we find the duration $s$ of the first On pulse, also as a function of $r$,
\begin{equation}
\label{s_for_complementary}
\tilde{s} =  \cos^{-1}\left(-\frac{r^2 \mp B \sqrt{B^2+r^4-1}}{B^2+r^4}\right),
\end{equation}
where
\begin{equation}
\label{B_r}
B=\frac{r^2 \cos \frac{\pi }{4n}}{\sqrt{\frac{1}{r^2+1}-\cos ^2\frac{\pi }{4n}}}.
\end{equation}
The $\mp$ sign denotes that there are two solutions for the duration of the first On pulse and it is easy to prove that these two solutions add up to the duration of an intermediate (full) On pulse. Thus, there are two candidate optimal pulse-sequences with the same duration, one where the first pulse is shorter than the last and another where the order is reversed. 

Note that in Eq. (\ref{t_on_complementary}) the argument of $\sin ^{-1}$ should be confined in the interval $[-1 \; 1]$. From this requirement we find the maximum value $r$ can take for a complementary candidate optimal pulse-sequence containing $n$ Off pulses to exist, 
\begin{equation}
\label{omega_n_switch}
    r^n_{max}=\tan \frac{\pi }{4n}.
\end{equation}
Also, in Eq. (\ref{s_for_complementary}) it should be $B^2+r^4-1\geq 0$, which leads to the condition that $r$ should be larger than the minimum value 
\begin{equation}
\label{omega_type_switch}
    r^n_{min}=\sin \frac{\pi }{4n}.
\end{equation}
Thus, a complementary candidate optimal pulse-sequence containing $n$ Off pulses exists for $r$ in the range
\begin{equation}
\label{c_range}
r^n_{min}\leq r < r^n_{max}.
\end{equation}

\subsection{Symmetric Case}

The total propagator for a candidate optimal pulse-sequence containing $n$ Off pulses and for the case where the first and the last On pulses have equal durations $s$ is

\begin{eqnarray}
\label{total_prop_compl}
    U &=& U_s  U_{\text{off}}  \underbrace {U_{\text{on}} U_{\text{off}}\cdots U_{\text{on}} U_{\text{off}} }_{n-1}  U_{s} \nonumber
    \\
   \Rightarrow U_{s^*}  U &=& \underbrace{U_{s^*}U_{s}}_{\text{on}} U_{\text{off}}  \underbrace {U_{\text{on}} U_{\text{off}}\cdots U_{\text{on}} U_{\text{off}} }_{n-1}  U_{s} \nonumber
    \\
   \Rightarrow U_{t-s}  U &=& \underbrace {U_{\text{on}} U_{\text{off}}\cdots U_{\text{on}} U_{\text{off}} }_{n}  U_{s} \nonumber
    \\
   \Rightarrow U_{t-s}  U &=& (-1)^{n}  e^{-\frac{i}{2} n \alpha \hat{\nu}\cdot \hat{\sigma}}  U_{s} \nonumber
    \\
   \Rightarrow U &=& (-1)^{n}  U_{s-t}  e^{-\frac{i}{2} n \alpha \hat{\nu}\cdot \hat{\sigma}}  U_{s},
\end{eqnarray}



where $\alpha, \hat{\nu}$ are given in Eq. (\ref{nu_yz_r}).
Using identity (\ref{identity}) we find that
\begin{equation}
\label{total_propagator_equal}
    U =  (-1)^{n}  e^{-\frac{i}{2} \beta_e \hat{\mu}_e\cdot \hat{\sigma}},
\end{equation}
where $\beta_e$
is the rotation angle and $\hat{\mu}_e = (\mu_{ex},\mu_{ey},\mu_{ez})^T$ is the rotation axis corresponding to the total propagator, with


\begin{widetext}


\begin{subequations}
\label{mu_xz}
    \begin{eqnarray}
    \beta_e &=& 2 \cos ^{-1}\left[\frac{ \sin \frac{n \alpha}{2} \cos \frac{\tilde{t}_{\text{on}}}{2} \sin \left(\tilde{s}-\frac{\tilde{t}_{\text{on}}}{2}\right)}{\sqrt{r^2+1} \sin \frac{\alpha }{2}}+\cos \frac{n \alpha}{2} \cos \left(\tilde{s}-\frac{\tilde{t}_{\text{on}}}{2}\right)\right], \label{be}
    \\[10pt]
    \mu_{ex} &=& \frac{r \left\{ 4 \sqrt{r^2+1} \cos \frac{n \alpha}{2} \sin \left(\tilde{s}-\frac{\tilde{t}_{\text{on}}}{2}\right)-\frac{2 \sin \frac{n \alpha}{2} \left[\cos \left(\tilde{s}-\tilde{t}_{\text{on}}\right)+\cos \tilde{s}-2\right]}{\sin \frac{\alpha }{2}}\right\}}{4 \left(r^2+1\right) \sin \frac{\beta }{2}}, \label{mex}
    \\[10pt]
    \mu_{ey} &=& 0,
    \\[10pt]
    \mu_{ez} &=& \frac{2 \sqrt{r^2+1} \cos \frac{n \alpha}{2} \sin \left(\tilde{s}-\frac{\tilde{t}_{\text{on}}}{2}\right) - \frac{\sin \frac{n \alpha}{2} \left[\cos \left(\tilde{s}-\tilde{t}_{\text{on}}\right)+\cos \tilde{s}+2 r^2\right]}{\sin \frac{\alpha }{2}}}{2 \left(r^2+1\right) \sin \frac{\beta }{2}}.  \label{mez} 
    \end{eqnarray}
\end{subequations}

\end{widetext}

\begin{figure}[h]
 \centering
 \includegraphics[width=\linewidth]{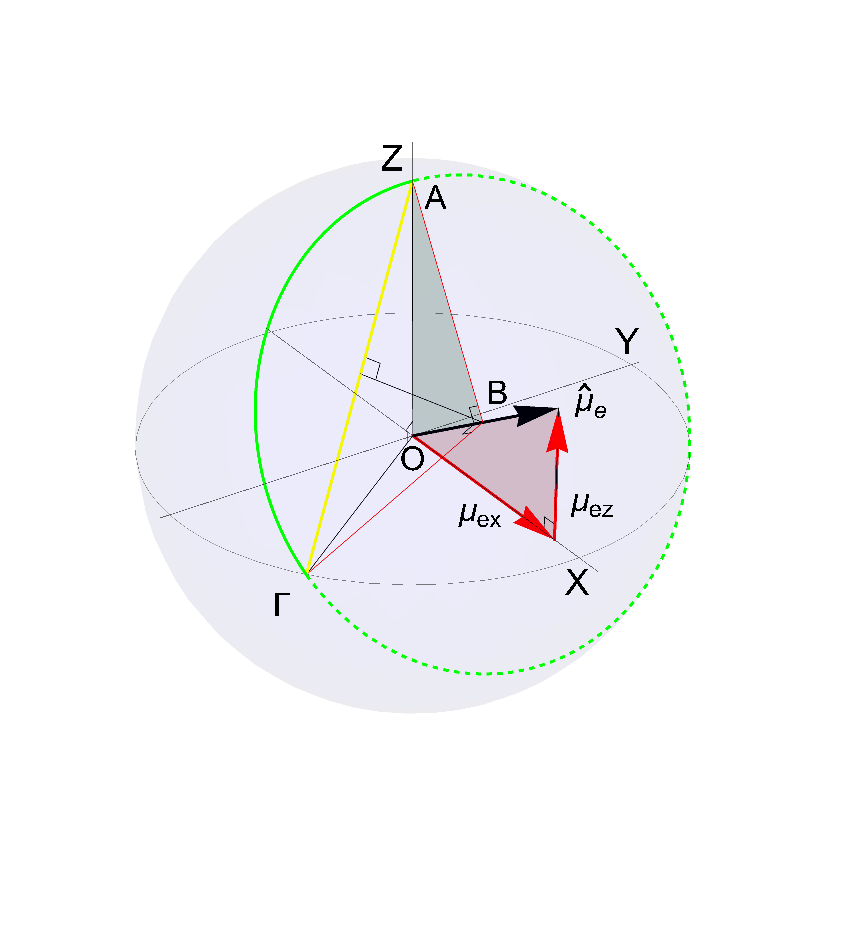}
\caption{Geometric interpretation of condition (\ref{condition_equal}).}
\label{fig:geom_sym}
\end{figure}

Observe that now the rotation axis is off the equatorial plane, on the $xz$-plane, thus the rotation angle required to reach the equator is larger than $\pi/2$.
Applying this rotation to the initial state we find the following probability amplitudes
of the final state 
\begin{eqnarray}
\label{c1_c0_final_equal}
    c_1(t_f) &=& (-1)^{n} \left( \cos \frac{\beta_e }{2}-i \sin \frac{\beta_e }{2} \mu_{ez} \right), \\
    c_0(t_f) &=& -i (-1)^{n} \sin \frac{\beta_e }{2} \mu_{ex}.
\end{eqnarray}
The final requirement $z(t_f)= \lvert c_1(t_f) \rvert^2 - \lvert c_0(t_f)\rvert^2$=0 leads to the condition
\begin{equation}
\label{condition_equal}
  2 \lvert\mu_{ex}\rvert \sin \frac{\beta_e}{2} =  \sqrt{2},
\end{equation}
from which we can calculate duration $s$ since both $\beta_e$ and $\mu_{ex}$ are functions of the variable $s$ and of parameters $\Omega_0,\Delta$. The geometric interpretation of Eq. (\ref{condition_equal}) is shown in Fig. \ref{fig:geom_sym}. The two shaded orthogonal triangles are equal, thus the radius AB of the circular arc (green solid line) traced by the tip of the Bloch vector is $\lvert\mu_{ex}\rvert$, while from the orthogonal triangle AO$\Gamma$ formed by the initial and final Bloch vectors we find the length of the corresponding chord A$\Gamma$ to be $\sqrt{2}$. But this length can be expressed in terms of the radius $\lvert\mu_{ex}\rvert$ and the corresponding central angle $\beta_e=\angle AB\Gamma$ as in the left hand side of Eq. (\ref{condition_equal}). The final state on the Bloch sphere is 
\begin{equation}
\label{c1_c0_final}
     \vec{r}(t_f) = \left( \frac{\mu _z}{\mu _x} , -\frac{\sqrt{\mu _x^2-\mu _z^2}}{\mu _x} , 0 \right).
\end{equation}

In order to find the range of parameter $r$ for which the symmetric candidate optimal solution with $n$ Off pulses exists, we use that the duration of the first and last On pulses should be restricted as $0<\tilde{s}<\tilde{t}_{\text{on}}$. Taking the upper limit $\tilde{s}\rightarrow \tilde{t}_{\text{on}}$, we easily find from Eq. (\ref{t_on(s)}) that both $\tilde{s}, \tilde{t}_{\text{on}}\rightarrow \pi$. Plugging these values in Eq. (\ref{be}) we immediately get $\beta_e=\pi$, while Eq. (\ref{alpha}) gives
\begin{equation}
\label{trig_a}
\cos{\frac{\alpha}{2}}=\frac{1}{\sqrt{r^2+1}},\quad \sin{\frac{\alpha}{2}}=\frac{r}{\sqrt{r^2+1}}.
\end{equation}
Using  successively Eqs. (\ref{condition_equal}), (\ref{mex}) and (\ref{trig_a}) we obtain
\begin{eqnarray}
    \mu_{ex} &=& \frac{1}{\sqrt{2}} \nonumber\\
    \frac{r}{\sqrt{r^2+1}}\cos \frac{n \alpha}{2}+\frac{1}{\sqrt{r^2+1}}\sin \frac{n \alpha}{2} &=& \frac{1}{\sqrt{2}} \nonumber\\
    \sin \frac{\alpha}{2}\cos \frac{n \alpha}{2}+\cos \frac{\alpha}{2}\sin \frac{n \alpha}{2} &=& \frac{1}{\sqrt{2}} \nonumber\\
    \sin \frac{(n+1)\alpha}{2} &=& \frac{1}{\sqrt{2}},
\end{eqnarray}
which leads to $\alpha=\pi/[2(n+1)]$. But from Eq. (\ref{trig_a}) we get
\begin{equation}
r=\tan\frac{\alpha}{2}=\tan\frac{\pi}{4(n+1)}=r^{n+1}_{max},
\end{equation}
and this is the limiting value of $r$ corresponding to $\tilde{s}\rightarrow \tilde{t}_{\text{on}}$. From Eq. (\ref{mez}) we find in this case that
\begin{equation}
\mu_{ez}=\cos{\frac{(n+1)\alpha}{2}}=\frac{1}{\sqrt{2}}.
\end{equation}
Working analogously for the lower limit $s\rightarrow 0$ we find $\tilde{t}_{\text{on}}\rightarrow \pi$, thus $\beta_e=\pi$ and Eq. (\ref{trig_a}) also holds. Then 
\begin{equation}
\mu_{ex} = \sin \frac{(n-1)\alpha}{2} = \frac{1}{\sqrt{2}}, \quad \mu_{ez} = -\cos \frac{(n-1)\alpha}{2} = -\frac{1}{\sqrt{2}},
\end{equation}
and $\alpha=\pi/[2(n-1)]$, for $n\neq 1$. From Eq. (\ref{trig_a}) we find the limiting value
\begin{equation}
r=\tan\frac{\alpha}{2}=\tan\frac{\pi}{4(n-1)}=r^{n-1}_{max},
\end{equation}
corresponding to  $\tilde{s}\rightarrow 0$. The symmetric candidate optimal solution with $n$ Off pulses, $n\geq 2$, exists for $r$ in the range
\begin{equation}
\label{e_range}
r^{n+1}_{max}\leq r< r^{n-1}_{max}.
\end{equation}
For $n=1$, i.e. the pulse-sequence On-Off-On with equal initial and final pulses, the limit $\tilde{s}\rightarrow 0$ is obviously not permitted. In this case we set as upper limit for $r$ the value $r^{0}_{max}=r^{1}_{max}=1$, above which a single On pulse is optimal.

\subsection{Synthesis of the two cases}

Now we compare the durations of the complementary and symmetric candidate optimal solutions which exist for each $r\leq 1$ in order to find the optimal. The total duration of the complementary pulse-sequence with n Off pulses is
\begin{eqnarray}
\label{T_c_n}
T_c^n\Delta &=& n\left(\pi+\frac{\tilde{t}_{\text{on}}}{\sqrt{r^2+1}}\right) \nonumber\\
      &=& n\left[\pi+\frac{2 \pi - 2\sin ^{-1}\left(\sqrt{r^2+1} \cos \frac{\pi }{4 n}\right)}{\sqrt{r^2+1}}\right],
\end{eqnarray}
where $r^{n}_{min}\leq r< r^{n}_{max}$. Next, we find the duration of the symmetric pulse-sequence with one Off pulse, $n=1$. Using Eq. (\ref{alpha}) in Eq. (\ref{mex}) we obtain
\begin{equation}
\mu_{ex}=\frac{r(1-\cos\tilde{s})}{(r^2+1)\sin\frac{\beta_e}{2}},
\end{equation}
condition (\ref{condition_equal}) leads to the solution
\begin{equation}
\tilde{s^*}=\cos^{-1}\left[1-\frac{1}{\sqrt{2}}\left(r+\frac{1}{r}\right)\right],
\end{equation}
thus
\begin{eqnarray}
  T_e^1\Delta &=& \pi+\frac{2\tilde{s}^*}{\sqrt{r^2+1}} \nonumber\\
              &=& \pi+\frac{2}{\sqrt{r^2+1}}\cos^{-1}\left[1-\frac{1}{\sqrt{2}}\left(r+\frac{1}{r}\right)\right].
\end{eqnarray}
In order to compare $T_e^1, T_c^1$ in the range $r^{1}_{min}=1/\sqrt{2}\leq r< r^{1}_{max}=r^{0}_{max}=1$, we form the function
\begin{eqnarray}
\label{g}
g(r) &=& \cos\left(\frac{T_c^1\Delta-\pi}{2}\sqrt{r^2+1}\right)-\cos\left(\frac{T_e^1\Delta-\pi}{2}\sqrt{r^2+1}\right) \nonumber\\
     &=& \frac{1}{\sqrt{2}}\left(r+\frac{1}{r}\right)-\sqrt{\frac{1-r^2}{2}}-1.
\end{eqnarray}
Since $g(1/\sqrt{2})=0$ and
\begin{equation}
g'(r)=\frac{r}{\sqrt{2(1-r^2)}}\left[1-\left(\frac{\sqrt{1-r^2}}{r}\right)^3\right]\geq 0,
\end{equation}
it is also $g(r)\geq 0$, which implies that $T_c^1\leq T_e^1$, since the cosines in Eq. (\ref{g}) are negative and the larger one corresponds to a smaller angle.

\begin{figure}[t]
 \centering
 \begin{subfigure}[b]{0.4\textwidth}
    \centering\caption{}\includegraphics[width=\linewidth]{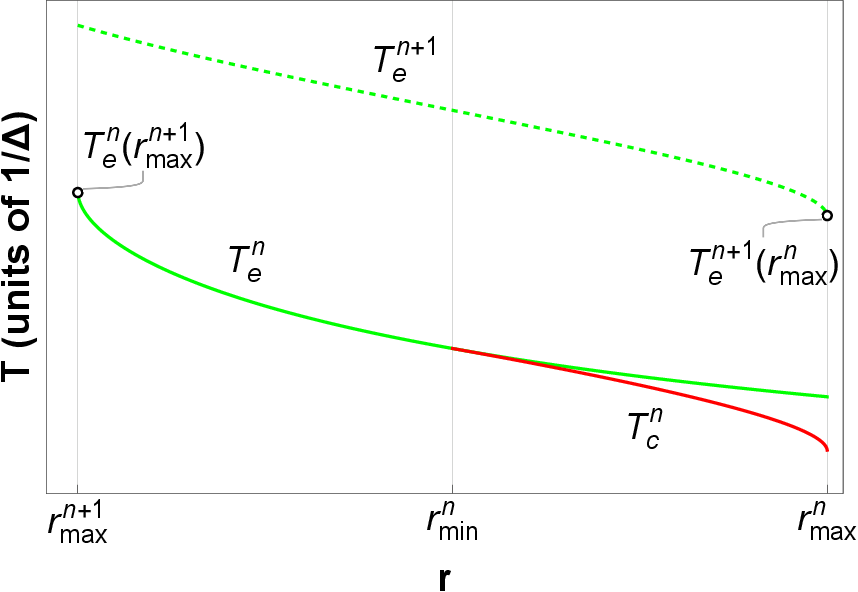}\label{fig:specific_area}
\end{subfigure}
\hspace{.2cm}
\begin{subfigure}[b]{0.4\textwidth}
    \centering\caption{}\includegraphics[width=\linewidth]{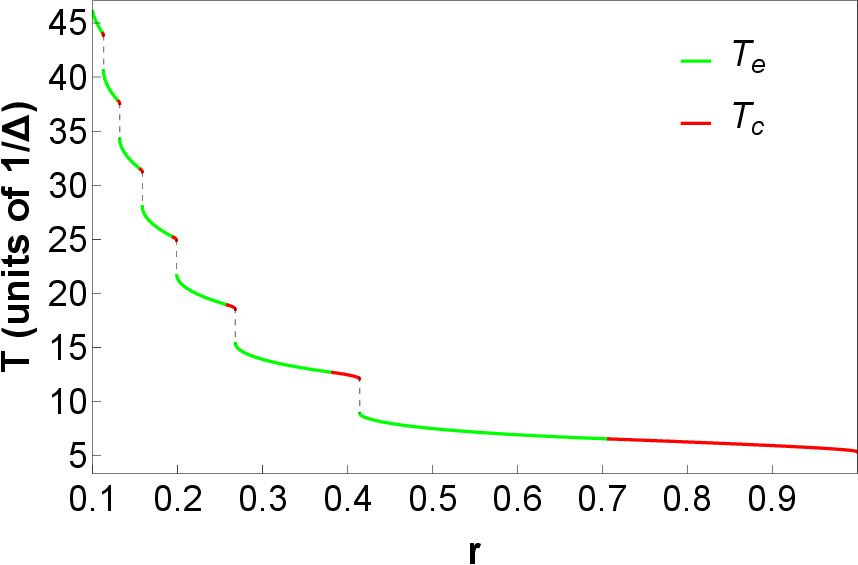}\label{fig:all_area}
\end{subfigure}
\caption{(a) Normalized durations of candidate optimal pulse-sequences when the dimensionless parameter $r=\Omega_0/\Delta$ is in the range $r^{n}_{min}\leq r< r^{n}_{max}$: symmetric pulse-sequence with $n$ Off pulses (green solid line), symmetric pulse-sequence with $n+1$ Off pulses (green dashed line), complementary pulse-sequence with $n$ Off pulses (red solid line). (b) Normalized duration of the optimal pulse sequence versus $r$, forming a ``stairway to heaven" as $r\rightarrow 0$.}
\label{fig:durations}
\end{figure}

For $n\geq 2$ the analytical calculation of $T_e^n$ from condition (\ref{condition_equal}) becomes cumbersome, if not impossible. For this reason, we recourse to the numerical solution of Eq. (\ref{condition_equal}) for each $r$ in the interval of interest, and express the duration of the pulse-sequence in terms of the solution $\tilde{s}^*$ as 
\begin{equation}
T_e^n\Delta=n\pi+\frac{2\tilde{s}^*+(n-1)\tilde{t}_{\text{on}}(\tilde{s}^*)}{\sqrt{r^2+1}}.
\end{equation}
The durations of the symmetric and complementary pulse-sequences in the interval $r^{n+1}_{max}\leq r< r^{n}_{max}$ are schematically depicted in Fig. \ref{fig:specific_area}
The comparison between the numerically obtained $T_e^n$ (green solid line) and the analytical $T_c^n$ (red solid line) reveals a simple pattern of optimality. For each interval $r^{n+1}_{max}\leq r< r^{n}_{max}$, the symmetric pulse-sequence with $n$ Off pulses is optimal in the segment $r^{n+1}_{max}\leq r< r^{n}_{min}$, while the complementary pulse-sequence with $n$ Off pulses is optimal in the segment $r^{n}_{min}\leq r< r^{n}_{max}$. This is consistent with the result for $n=1$ that was proved analytically above. We can intuitively understand the optimality of the complementary solution by observing that in this case the total rotation angle is $\beta_c=\pi/2$, while in the symmetric case from condition (\ref{condition_equal}) we find that $\sin{(\beta_e/2)}=1/(\mu_{ex}\sqrt{2})\geq 1/\sqrt{2}$, thus $\beta_e\geq\pi/2$. Note that for $r=r^{n}_{min}$ it is $T_c^n=T_e^n$. Indeed, it can be easily shown that this point corresponds to the solution $\tilde{s}^*=\tilde{t}_{\text{on}}(\tilde{s}^*)/2$, thus the complementary and symmetric sequences coincide.

Another interesting observation is that in the interval under consideration, a candidate optimal pulse-sequence is also the symmetric with $n+1$ Off pulses (green dashed line), but it always behaves worse than the symmetric sequence with $n$ Off pulses. We can understand this behavior as follows. First note that obviously $T_e^n$ is a decreasing function of $r$, since for a larger control bound $r$ the Bloch vector can be brought on the equator at least in the timed needed with smaller $r$. Next, we find and compare the values $T_e^n(r^{n+1}_{max})$ and $T_e^{n+1}(r), r\rightarrow r^{n}_{max}$, i.e the maximum value of $T_e^n$ on the left and the minimum value approached by $T_e^{n+1}$ on the right of the interval, see Fig. \ref{fig:specific_area}. From the analysis of the previous subsection we have that in the first case $\tilde{s}*, \tilde{t}_{\text{on}}\rightarrow \pi$, while in the second case $\tilde{s}^*\rightarrow 0$, $\tilde{t}_{\text{on}}\rightarrow \pi$, thus
\begin{eqnarray}
T_e^n(r^{n+1}_{max})\Delta &=& n\pi+\frac{(n+1)\pi}{\sqrt{(r^{n+1}_{max})^2+1}}, \\
T_e^{n+1}(r)\Delta &\xrightarrow[r \to r^{n}_{max}]{} & (n+1)\pi+\frac{n\pi}{\sqrt{(r^{n}_{max})^2+1}}.
\end{eqnarray}
In the large $n$ limit it is $r^{n}_{max}\rightarrow 0$ and the above expressions for the maximum and minimum durations of the symmetric $n$ and $n+1$ branches, respectively, tend to the common value $(2n+1)\pi$.

Putting all this information together, in Fig. \ref{fig:all_area} we plot the duration of the optimal pulse-sequence versus $r$, with the corresponding color (green or red) indicating whether the sequence is of the symmetric or complementary form, respectively. Observe that a ``stairway to heaven" diagram is formed. 

\begin{figure}[t]
 \centering
 \begin{subfigure}[b]{0.23\textwidth}
    \centering\caption{}\includegraphics[width=\linewidth]{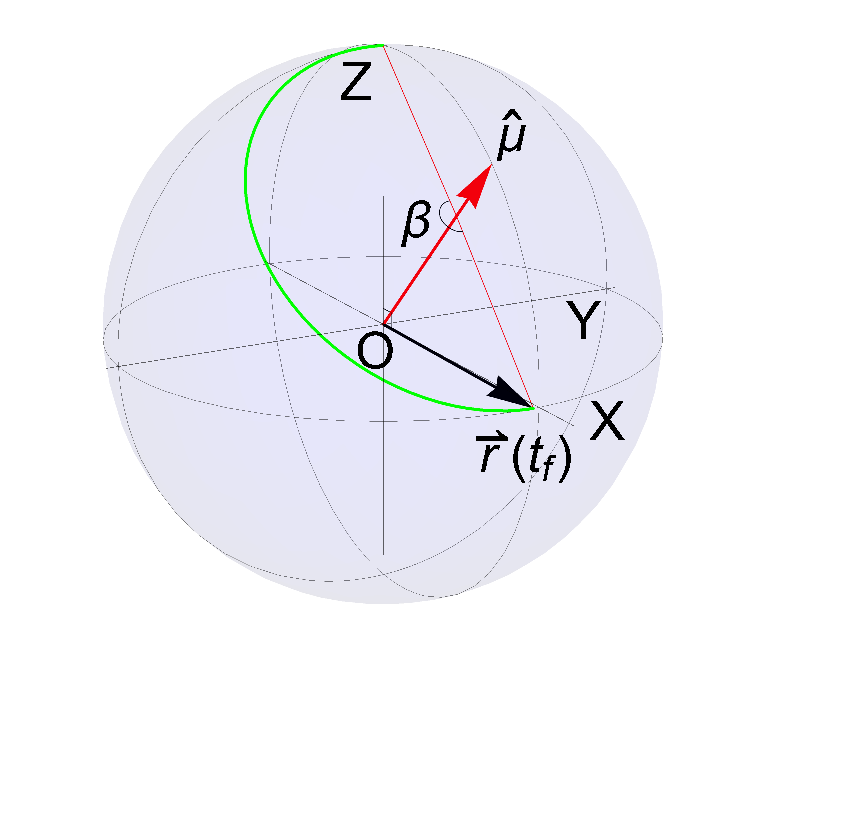}\label{fig:final_state(a)}
\end{subfigure}
\begin{subfigure}[b]{0.23\textwidth}
    \centering\caption{}\includegraphics[width=\linewidth]{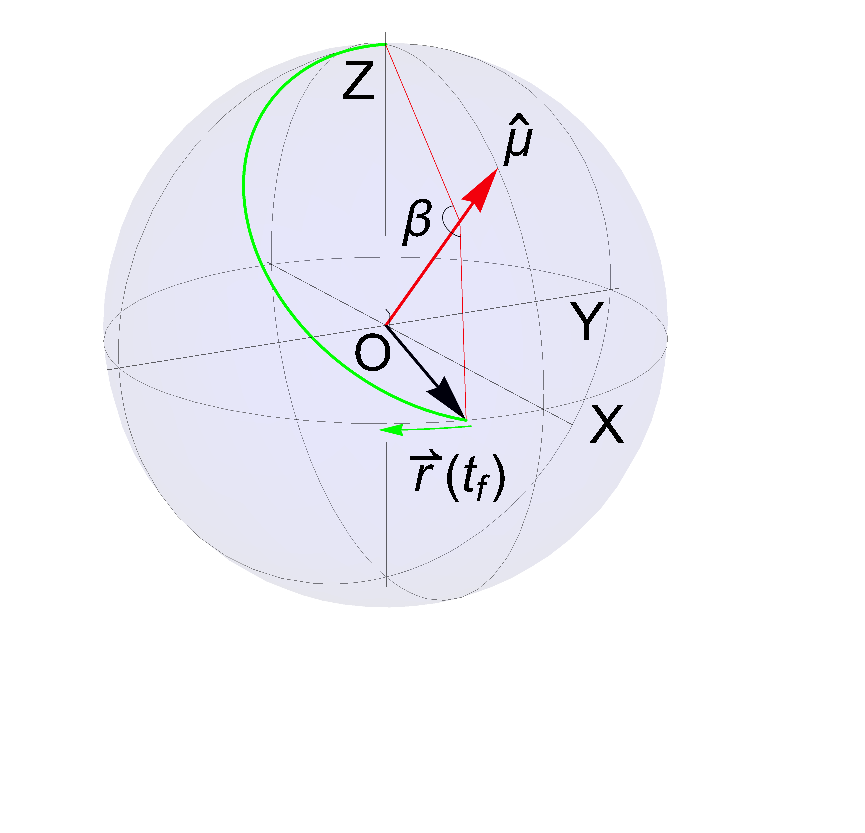}\label{fig:final_state(b)}
\end{subfigure} \\
\begin{subfigure}[b]{0.23\textwidth}
    \centering\caption{}\includegraphics[width=\linewidth]{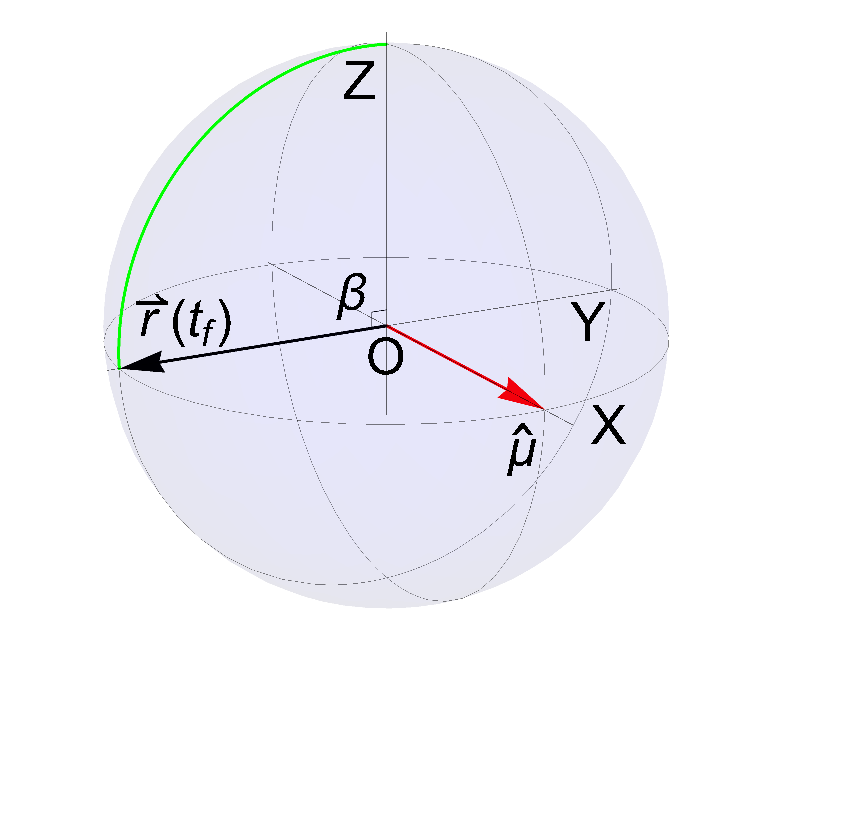}\label{fig:final_state(c)}
\end{subfigure}
\begin{subfigure}[b]{0.23\textwidth}
    \centering\caption{}\includegraphics[width=\linewidth]{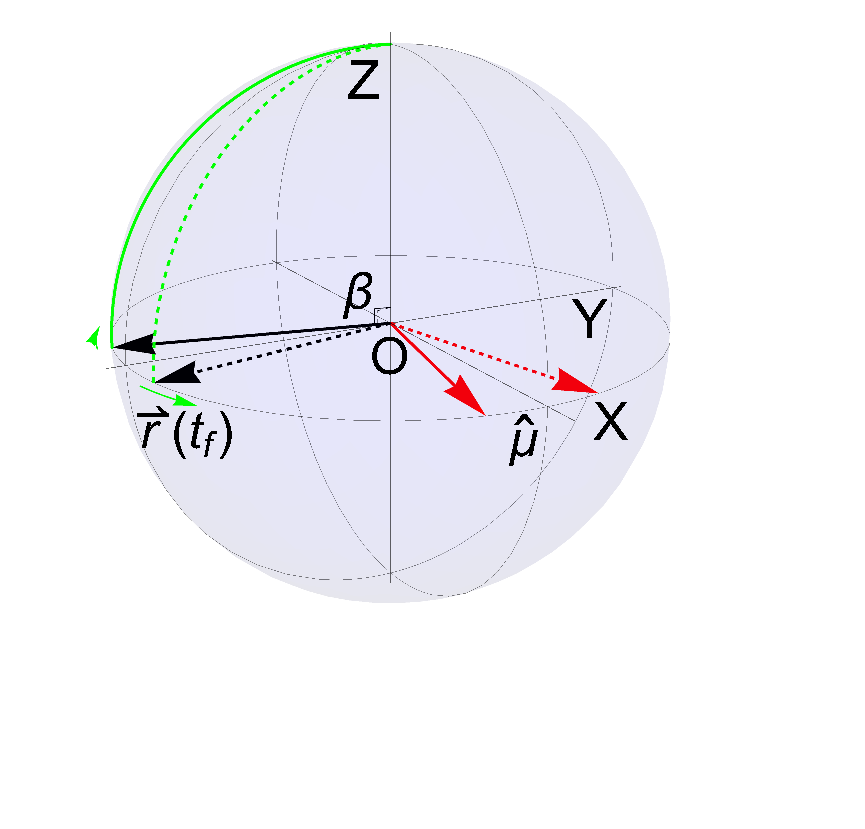}\label{fig:final_state(d_long)}
\end{subfigure}
\caption{Final Bloch vector on the equator $\vec{r}(t_f)$ (black arrow), as well as rotation angle $\beta$ and axis $\hat{\mu}$ (red arrow) corresponding to the total propagator of the optimal sequence, for various values of parameter $r=\Omega_0/\Delta$ in the interval $r^{n+1}_{max}\leq r< r^{n}_{max}$: (a) $r=r^{n+1}_{max}$, (b) $r^{n+1}_{max}< r< r^{n}_{min}$, (c) $r = r^{n}_{min}$, (d) $r^{n}_{min}< r< r^{n}_{max}$.}
\label{fig:final_state}
\end{figure}

In Fig. \ref{fig:final_state} we display the final Bloch vector on the equator $\vec{r}(t_f)$, as well as the rotation angle $\beta$ and axis $\hat{\mu}$ corresponding to the total propagator of the optimal sequence, for various values of parameter $r$ in the interval $r^{n+1}_{max}\leq r< r^{n}_{max}$. For $r=r^{n+1}_{max}$ the optimal pulse sequence is the symmetric one, with $\beta_e=\pi$ and $\hat{\mu}_e=(1/\sqrt{2},0,1/\sqrt{2})^T$, 
thus $\vec{r}(t_f)=(1,0,0)^T$, see Fig. \ref{fig:final_state(a)}. For $r^{n+1}_{max}< r< r^{n}_{min}$ the symmetric pulse-sequence remains optimal, but $\mu_{ex}$ increases and $\beta_e$ decreases, while $\vec{r}(t_f)$ is rotated clockwise, as displayed in Fig. \ref{fig:final_state(b)}. For $r = r^{n}_{min}$ the symmetric and complementary sequences coincide, $\beta=\pi/2$ and $\hat{\mu}=(1,0,0)^T$, thus $\vec{r}(t_f)=(0,-1,0)^T$, see Fig. \ref{fig:final_state(c)}. For $r^{n}_{min}< r< r^{n}_{max}$ the complementary pulse sequence becomes optimal with $\beta_c=\pi/2$, see Fig. \ref{fig:final_state(d_long)}. For the complementary solution starting with the shorter On pulse, as $r$ increases towards $r^{n}_{max}$, vector $\hat{\mu}_c$ (dashed red arrow) is rotated towards $(0,1,0)^T$ counterclockwise, thus $\vec{r}(t_f)$ (dashed black arrow) is rotated in the same sense towards $(1,0,0)^T$. In the case of the complementary solution starting with the longer On pulse, $\hat{\mu}_c$ (solid red arrow) is rotated clockwise towards $(0,-1,0)^T$, while $\vec{r}(t_f)$ (solid black arrow) is rotated in the same sense towards $(-1,0,0)^T$.

\subsection{An intuitive sub-optimal control}

\begin{figure}[t]
 \centering
 \includegraphics[width=\linewidth]{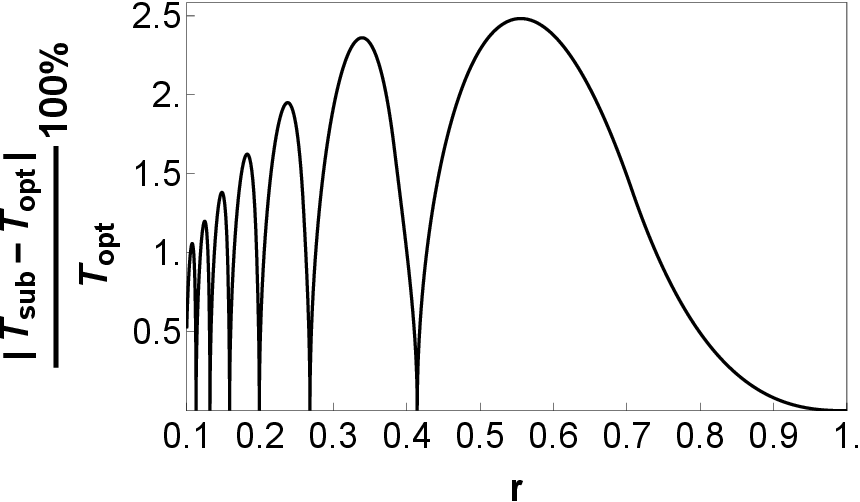}
\caption{Percentage deviation between the durations of the intuitive sub-optimal and optimal pulse-sequences.}
\label{fig:percentage_deviation}
\end{figure}

Motivated by the form of the optimal control, we consider a simpler suboptimal pulse-sequence where each On pulse except the first has duration $\tilde{t}_{on}=\pi$, while the duration $s$ of the first On pulse is determined from the final requirement to bring the Bloch vector on the equator. The total propagator is now
\begin{eqnarray}
\label{total_prop_suboptimal}
    U &=& \underbrace {U_{\text{on}} U_{\text{off}}\cdots U_{\text{on}} U_{\text{off}} }_{n}  U_{s} \nonumber\\
      &=& (U_{\text{on}} U_{\text{off}})^n U_{s} \nonumber\\
      &=& (-1)^n  e^{-\frac{i}{2} n \alpha \sigma_y} e^{-\frac{i}{2} \tilde{s} \hat{n} \cdot \hat{\sigma}},
\end{eqnarray}
where from Eq. (\ref{alpha})
\begin{equation}
\label{sub_alpha}
\alpha = 2\tan ^{-1}r,
\end{equation}
while $\hat{n}=(\sin{\frac{\alpha}{2}}, 0, \cos{\frac{\alpha}{2}})^T$ from Eq. (\ref{propagator_on_general}).
Applying this propagator to the initial state and demanding the final Bloch vector to lie on the equator we get
\begin{equation}
\label{s_suboptimal}
    \cos \tilde{s} = \frac{\cos [(n+1) \alpha]+\cos n\alpha}{\cos [(n+1) \alpha]-\cos n\alpha}.
\end{equation}
Observe from Eq. (\ref{sub_alpha}) that for $r^{n+1}_{max}\leq r< r^{n}_{max}$ it is $\pi/[2(n+1)]\leq \alpha <\pi/(2n)$. For $\alpha=\pi/[2(n+1)]$ Eq. (\ref{s_suboptimal}) gives $\tilde{s}=\pi$, while in the limit $\alpha\rightarrow\pi/(2n)$ it gives $\tilde{s}\rightarrow 0$. In Fig. \ref{fig:percentage_deviation} we plot the percentage deviation between the durations of the sub-optimal and optimal pulse-sequences, as a function of parameter $r$. Observe that it does not exceed $2.5\%$.

\section{Examples}
\label{sec:results}

\begin{figure*}[t]
 \centering
 \vspace{-1.5cm}
 \begin{subfigure}[t!]{0.4\textwidth}
    \centering\caption{}\vspace{1.1cm}\includegraphics[width=\linewidth]{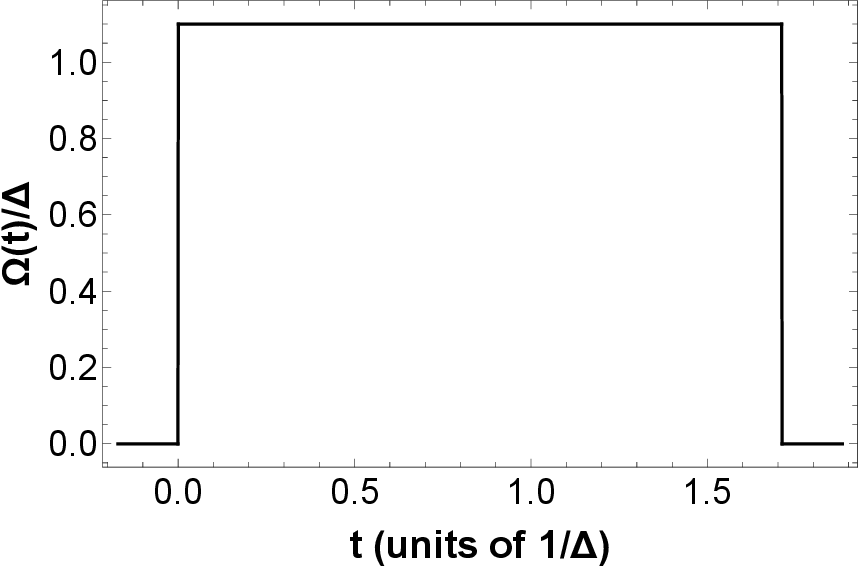}\label{fig:r_1_1_control}
\end{subfigure}
\begin{subfigure}[t!]{0.4\textwidth}
    \centering\vspace{1.4cm}\caption{}\includegraphics[width=\linewidth]{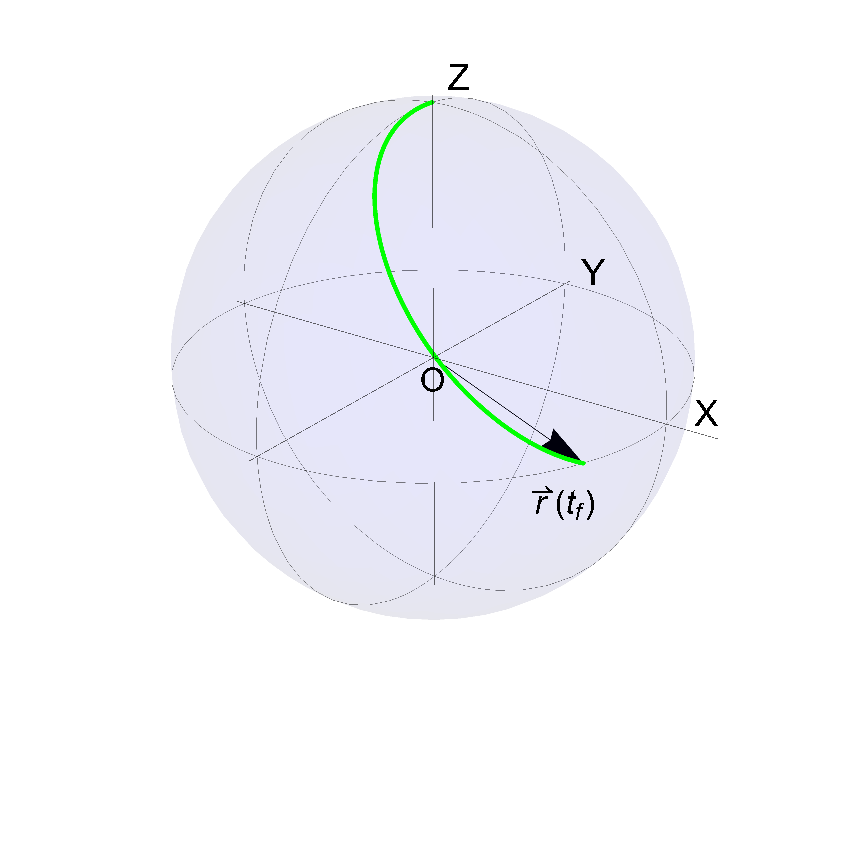}\label{fig:r_1_1_sphere}
\end{subfigure} \\
\vspace{-2.5cm} 
\begin{subfigure}[c]{0.4\textwidth}
    \centering\caption{}\vspace{1.1cm}\includegraphics[width=\linewidth]{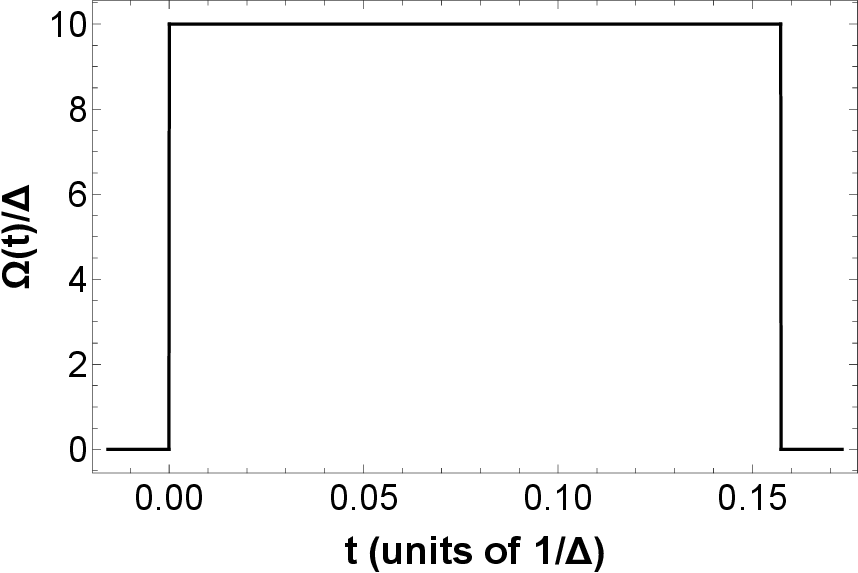}\label{fig:r_10_control}
\end{subfigure}
\begin{subfigure}[c]{0.4\textwidth}
    \centering\vspace{1.4cm}\caption{}\includegraphics[width=\linewidth]{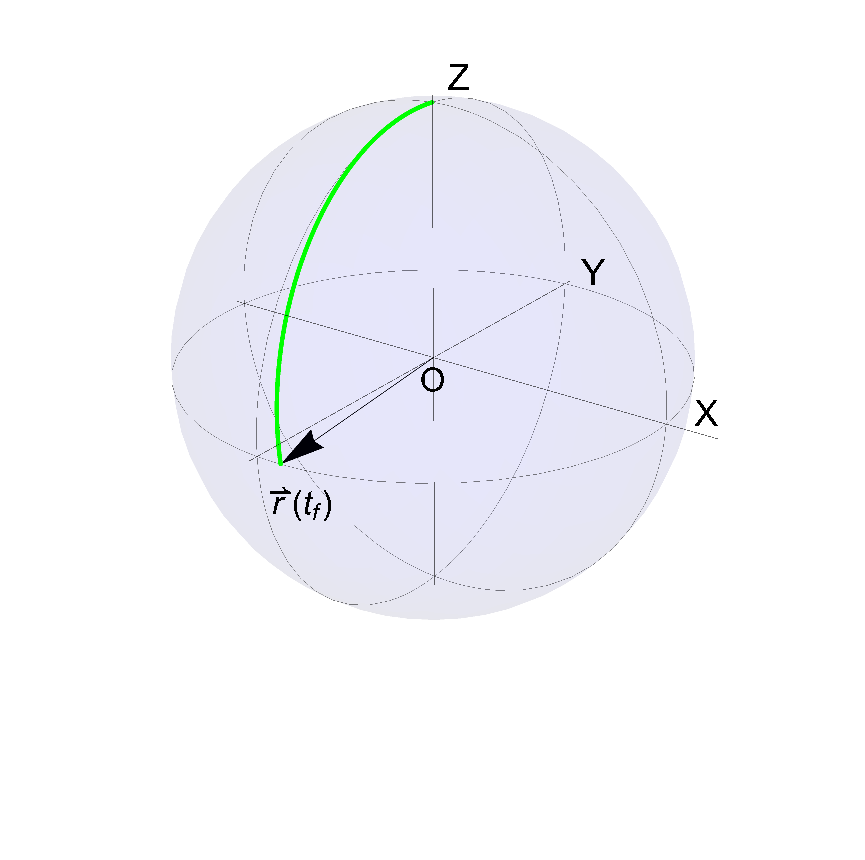}\label{fig:r_10_sphere}
\end{subfigure}
\caption{Optimal pulse-sequences containing a single On pulse: (a) Optimal pulse for $r=1.1$. (b) Corresponding optimal trajectory. (c) Optimal pulse for $r=10$. (d) Corresponding optimal trajectory.}
\label{fig:final_state_r_g_1}
\end{figure*}

\begin{figure*}[t]
 \centering
 \vspace{-1.5cm}
 \begin{subfigure}[t!]{0.4\textwidth}
    \centering\caption{}\vspace{1.1cm}\includegraphics[width=\linewidth]{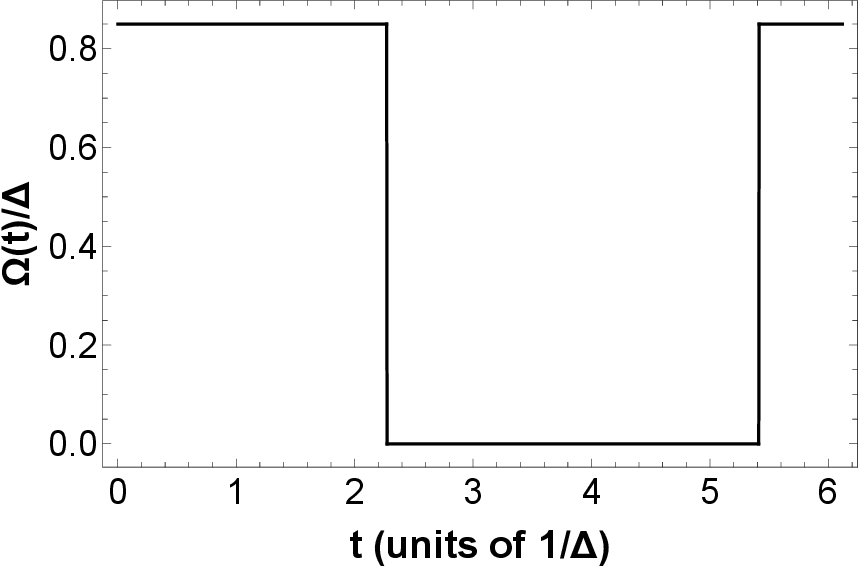}\label{fig:n_1_com_control}
\end{subfigure}
\begin{subfigure}[t!]{0.4\textwidth}
    \centering\vspace{1.4cm}\caption{}\includegraphics[width=\linewidth]{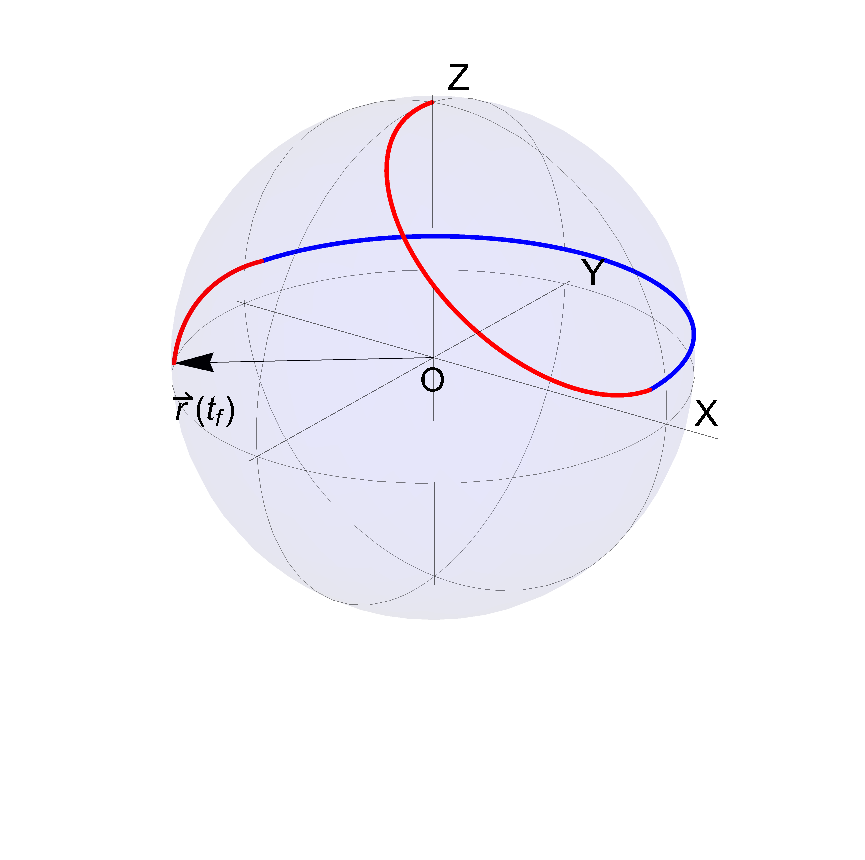}\label{fig:n_1_com_sphere}
\end{subfigure} \\
\vspace{-2.5cm} 
\begin{subfigure}[c]{0.4\textwidth}
    \centering\caption{}\vspace{1.1cm}\includegraphics[width=\linewidth]{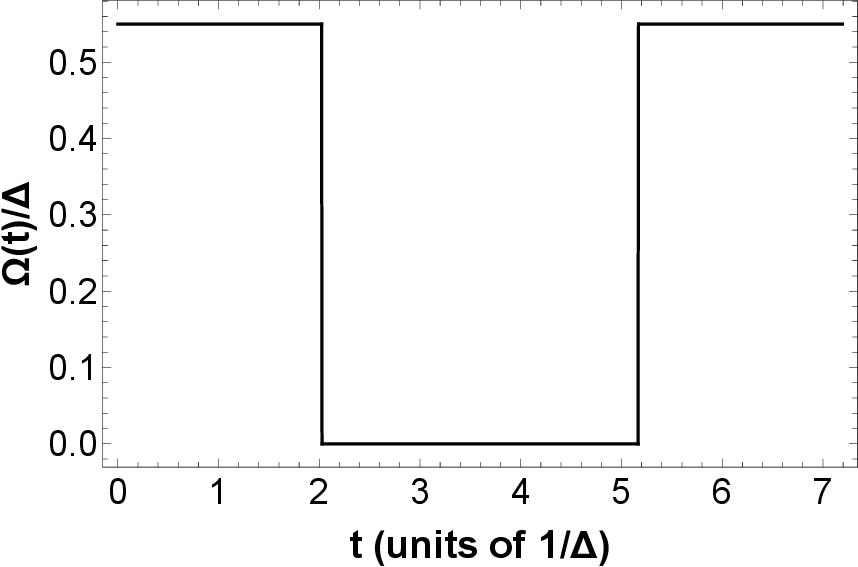}\label{fig:n_1_sym_control}
\end{subfigure}
\begin{subfigure}[c]{0.4\textwidth}
    \centering\vspace{1.4cm}\caption{}\includegraphics[width=\linewidth]{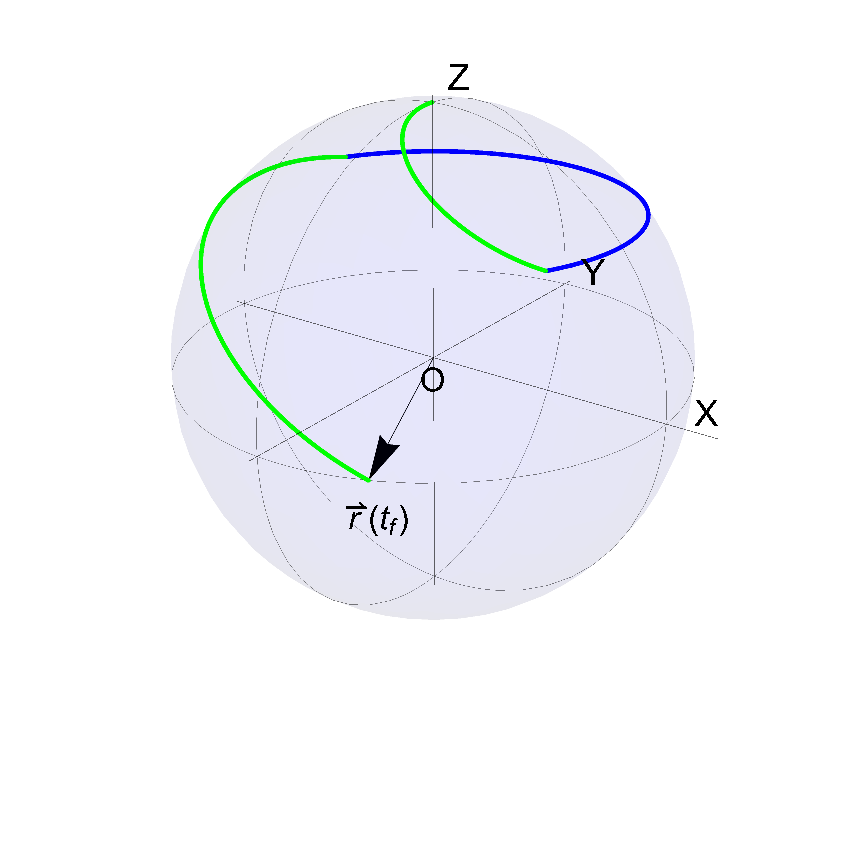}\label{fig:n_1_sym_sphere}
\end{subfigure}
\caption{Examples of optimal pulse-sequences containing a single Off pulse: (a) Optimal complementary pulse-sequence for $r=0.85$. (b) Corresponding optimal trajectory, where the red segments correspond to On pulses while the blue segment to the Off pulse. (c) Optimal symmetric pulse-sequence for $r=0.55$. (d) Corresponding optimal trajectory, where the green segments correspond to On pulses while the blue segment to the Off pulse.}
\label{fig:final_state_n_1}
\end{figure*}

\begin{figure*}[t]
 \centering
 \vspace{-1.5cm}
 \begin{subfigure}[t!]{0.4\textwidth}
    \centering\caption{}\vspace{1.1cm}\includegraphics[width=\linewidth]{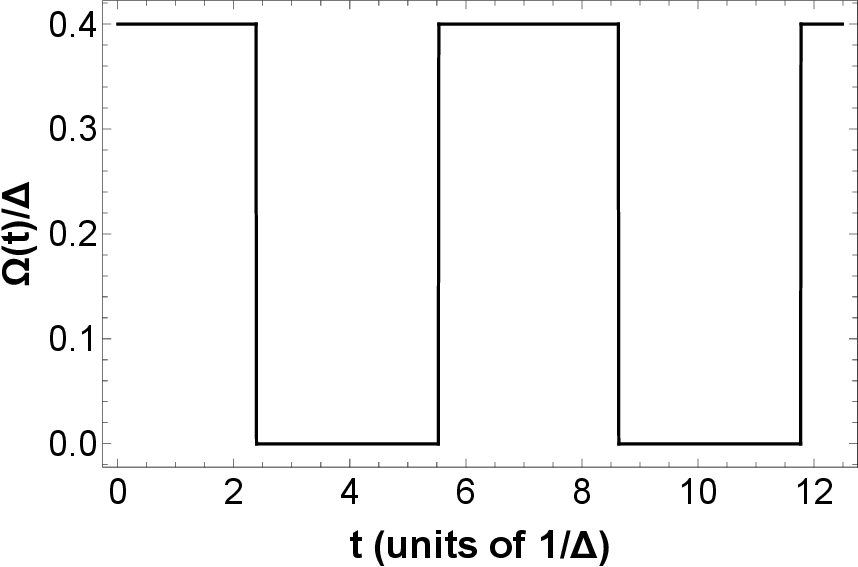}\label{fig:n_2_com_control}
\end{subfigure}
\begin{subfigure}[t!]{0.4\textwidth}
    \centering\vspace{1.4cm}\caption{}\includegraphics[width=\linewidth]{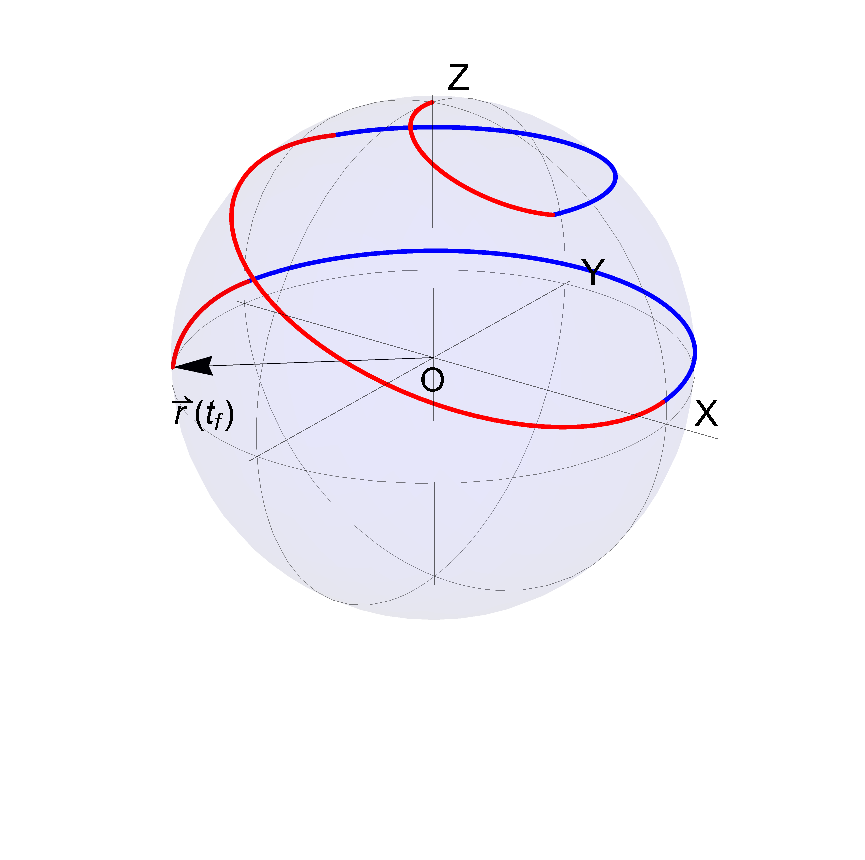}\label{fig:n_2_com_sphere}
\end{subfigure} \\
\vspace{-2.5cm} 
\begin{subfigure}[c]{0.4\textwidth}
    \centering\caption{}\vspace{1.1cm}\includegraphics[width=\linewidth]{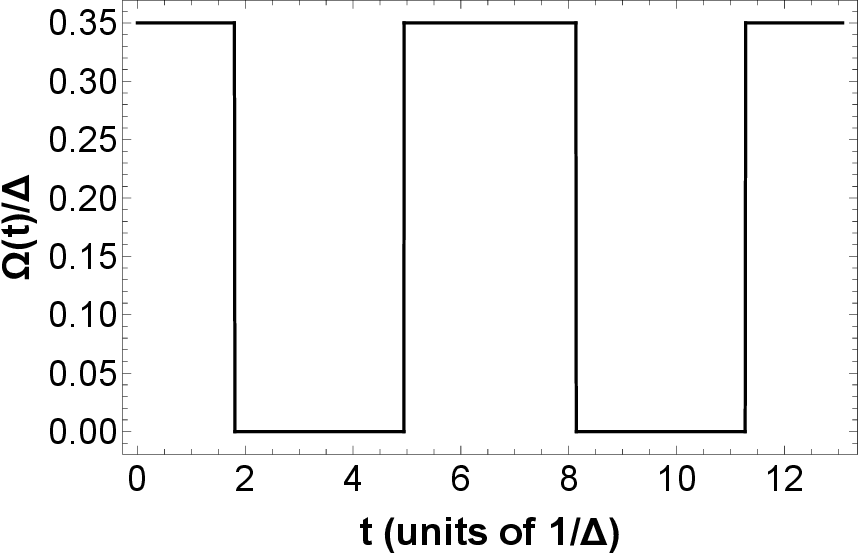}\label{fig:n_2_sym_control}
\end{subfigure}
\begin{subfigure}[c]{0.4\textwidth}
    \centering\vspace{1.4cm}\caption{}\includegraphics[width=\linewidth]{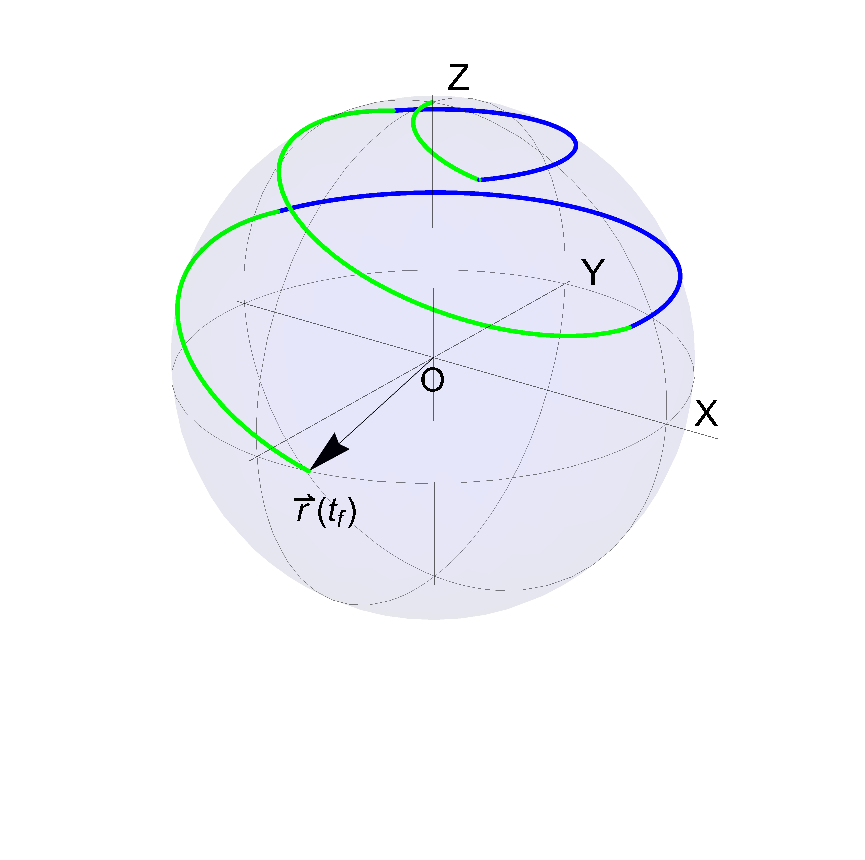}\label{fig:n_2_sym_sphere}
\end{subfigure}
\caption{Examples of optimal pulse-sequences containing two Off pulses: (a) Optimal complementary pulse-sequence for $r=0.4$. (b) Corresponding optimal trajectory, where the red segments correspond to On pulses while the blue segments to Off pulses. (c) Optimal symmetric pulse-sequence for $r=0.35$. (d) Corresponding optimal trajectory, where the green segments correspond to On pulses while the blue segments to Off pulses.}
\label{fig:final_state_n_2}
\end{figure*}

\begin{figure*}[t]
 \centering
 \vspace{-1.5cm}
 \begin{subfigure}[t!]{0.4\textwidth}
    \centering\caption{}\vspace{1.1cm}\includegraphics[width=\linewidth]{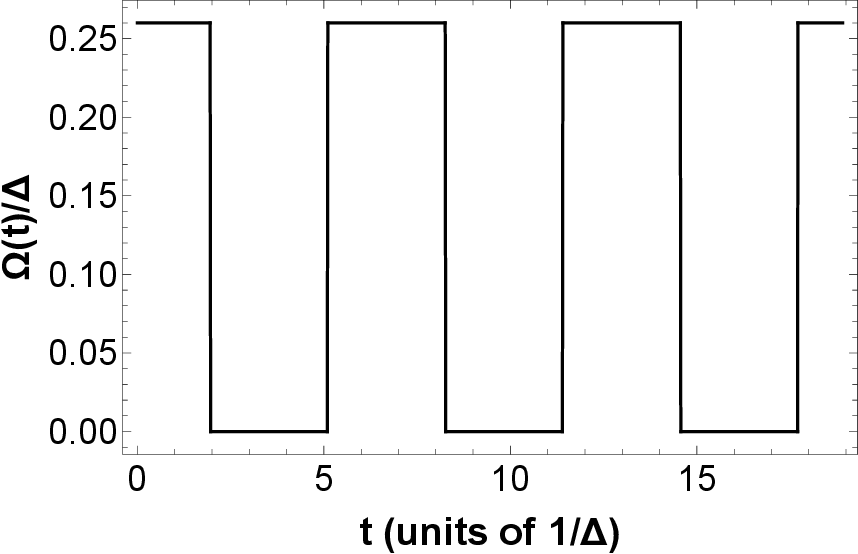}\label{fig:n_3_com_control}
\end{subfigure}
\begin{subfigure}[t!]{0.4\textwidth}
    \centering\vspace{1.4cm}\caption{}\includegraphics[width=\linewidth]{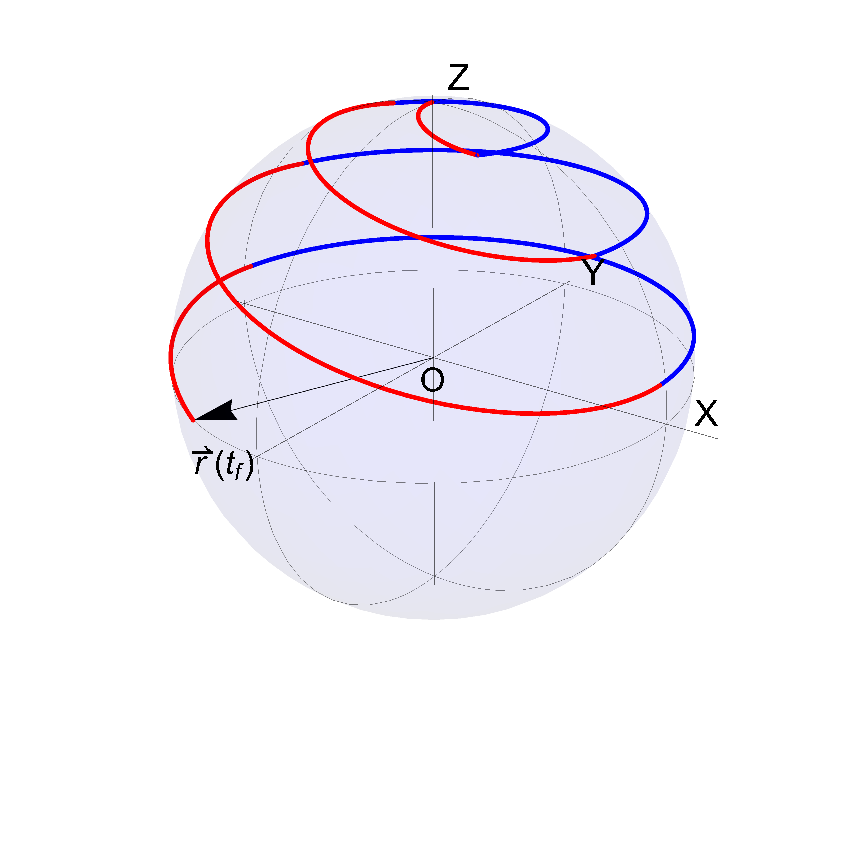}\label{fig:n_3_com_sphere}
\end{subfigure} \\
\vspace{-2.5cm} 
\begin{subfigure}[c]{0.4\textwidth}
    \centering\caption{}\vspace{1.1cm}\includegraphics[width=\linewidth]{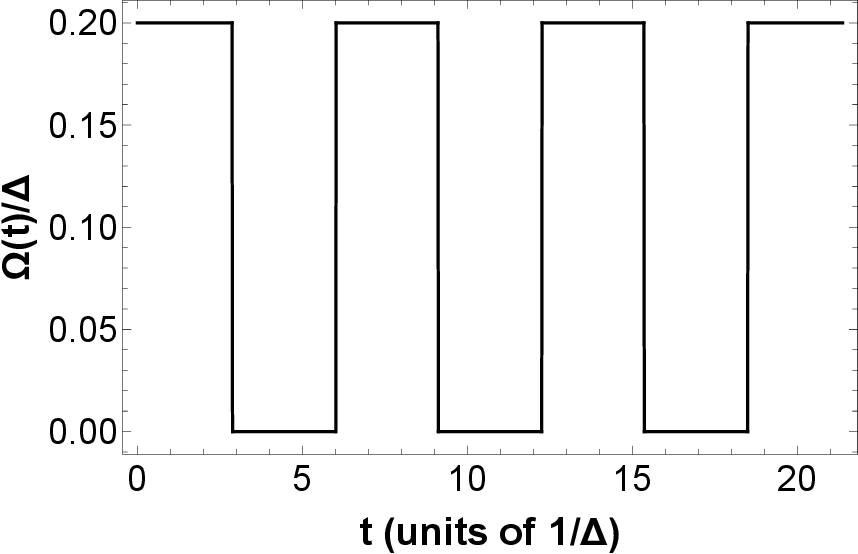}\label{fig:n_3_sym_control}
\end{subfigure}
\begin{subfigure}[c]{0.4\textwidth}
    \centering\vspace{1.4cm}\caption{}\includegraphics[width=\linewidth]{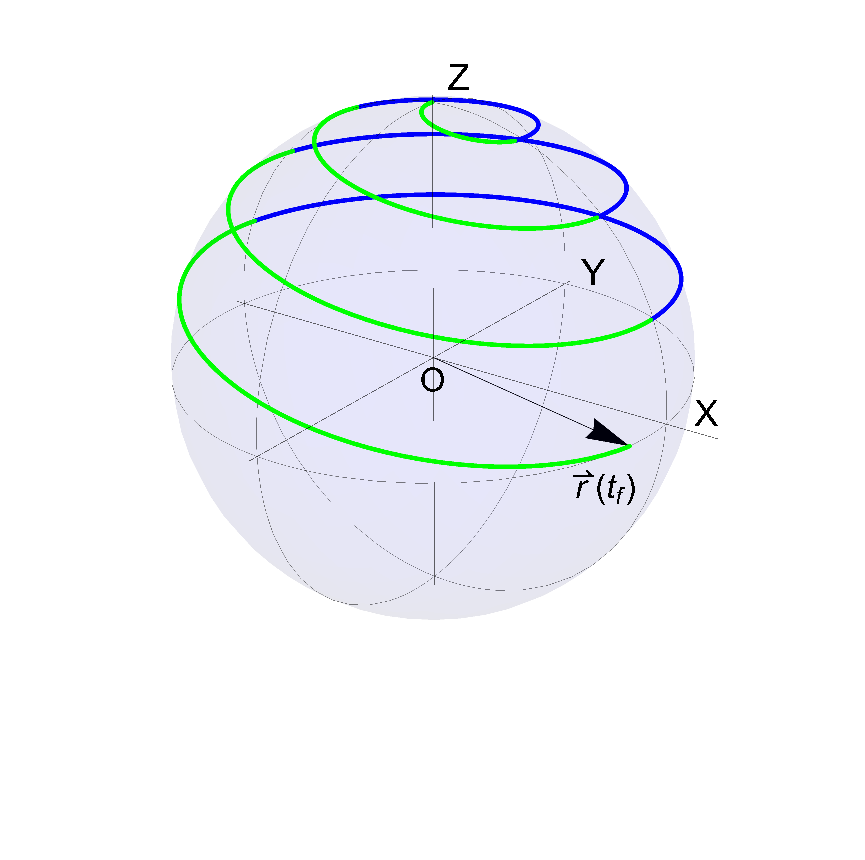}\label{fig:n_3_sym_sphere}
\end{subfigure}
\caption{Examples of optimal pulse-sequences containing three Off pulses: (a) Optimal complementary pulse-sequence for $r=0.26$. (b) Corresponding optimal trajectory, where the red segments correspond to On pulses while the blue segments to Off pulses. (c) Optimal symmetric pulse-sequence for $r=0.2$. (d) Corresponding optimal trajectory, where the green segments correspond to On pulses while the blue segments to Off pulses.}
\label{fig:final_state_n_3}
\end{figure*}

In this section we present examples of optimal controls and trajectories for various values of parameter $r$. We start with values $r\geq 1$, for which the optimal control is a single On pulse. In Fig. \ref{fig:r_1_1_control} we show the optimal pulse for $r=1.1$ and in Fig. \ref{fig:r_1_1_sphere} the corresponding optimal trajectory. Similar results are displayed in Figs. \ref{fig:r_10_control} and \ref{fig:r_10_sphere}
for the value $r=10$. Observe that, as $r$ increases, the optimal pulse tends to a $\pi/2$ pulse. Next, we consider values of $r$ for which the optimal pulse-sequence contains one Off pulse. In Fig. \ref{fig:n_1_com_control} we show the optimal sequence for $r=0.85$, which is of the complementary type, while in Fig. \ref{fig:n_1_com_sphere} we plot the corresponding optimal trajectory. Note that the red segments correspond to On pulses while the blue segment to the Off pulse. In Fig. \ref{fig:n_1_sym_control} we show the optimal sequence for $r=0.55$, which is of the symmetric type, while in Fig. \ref{fig:n_1_sym_sphere} we plot the corresponding optimal trajectory. Now the green segments correspond to On pulses, in consistency with the colors used in Fig. \ref{fig:durations}, while the blue segment to the Off pulse. Finally, in Figs. \ref{fig:final_state_n_2}, \ref{fig:final_state_n_3} we present examples of complementary and symmetric optimal pulse-sequences with two and three Off pulses, respectively.

\section{Conclusion and future work}
\label{sec:conclusion}

In this article, we used Pontryagin's Maximum Principle to solve the problem of generating in minimum time a uniform superposition of states in a two-level system, with fixed energy spacing (detuning) between the two levels and a single transverse control field restricted between zero and a maximum amplitude. For each value of the ratio between the maximum control amplitude and the detuning we derived the optimal pulse-sequence and calculated the durations of the corresponding pulses. The presented framework can be also exploited in the problem of fast charging a quantum battery based on a two-level system, as well as for the optimization of pulse-sequences for the controlled preparation of the excited state in a quantum emitter, which is a prerequisite for its usage as a single-photon source. Additionally, it can be extended for the efficient quantum control of interacting qubit arrays, especially for the cases where only global transverse control fields are employed as in Ref. \cite{Stojanovic22b}, and in combination with numerical optimal control for arrays with many qubits. 

\begin{acknowledgements}
The present work was financially supported by the ``Andreas Mentzelopoulos Foundation". The work of D.S. was funded by an Empirikion Foundation research grant.
\end{acknowledgements}



\bibliographystyle{apsrev4-2}
\bibliography{main}

\vspace*{1. cm}

\end{document}